\documentclass[12pt]{article}      
\usepackage{amsmath,amssymb,amsfonts} 
\usepackage[hypertex]{hyperref}      
\usepackage[dvips]{graphicx}              

\numberwithin{equation}{section} 

\newcommand{\N}{{\cal{N}}}
\newcommand{\half}{\frac{1}{2}}
\newcommand{\be}{\begin{equation}}
\newcommand{\ee}{\end{equation}} 
\newcommand{\ba}{\begin{eqnarray}}
\newcommand{\ea}{\end{eqnarray}}
\newcommand{\nn}{\nonumber}
\newcommand{\tr}{ {\rm{Tr} }}

\newcommand{\cO}{{\cal{O}}}
\newcommand{\ab}{ \vert Z^1 \vert^2 } 
\newcommand{\bPhi}{\bar{ \Phi} }
\newcommand{\CS}{{\cal{S}}}

\newcommand{\Rb}{\mathbb{R}}
\newcommand{\Cb}{\mathbb{C}}
\newcommand{\Zb}{\mathbb{Z}}
\DeclareMathOperator{\diag}{\mathrm{diag}}
\newcommand{\del}{\partial}
\newcommand{\delb}{\bar{\partial}}

\newcommand{\TD}{T_{D3}}

\begin{document}

\thispagestyle{empty}

 \renewcommand{\thefootnote}{\fnsymbol{footnote}}
\begin{flushright}
 \begin{tabular}{l}
 {\tt arXiv:0812.1420[hep-th]}\\
 {SNUTP 08-011}
 \end{tabular}
\end{flushright}

 \vfill
 \begin{center}
 {\bfseries \Large Holography of BPS surface operators}
\vskip 1.9 truecm

\noindent{{\large
  Eunkyung Koh \footnote{ekoh(at)phya.snu.ac.kr}
and Satoshi Yamaguchi \footnote{yamaguch(at)phya.snu.ac.kr} }}
\bigskip
 \vskip .9 truecm
\centerline{\it Department of Physics and Astronomy,
Seoul National University,
Seoul 151-747, KOREA}
\vskip .4 truecm
\end{center}
 \vfill
\vskip 0.5 truecm

\begin{abstract}
We study a class of dilatation invariant BPS surface operators in 4-dimensional $\N=4$ Super Yang-Mills theory and their holographic duals in type IIB string theory in $AdS_5\times S^5$.  First we take an example of 1/4 BPS surface operator and study it in detail from the holographic point of view. The gravity dual of this surface operator is a D3-brane characterized by a holomorphic submanifold. The supersymmetry and vacuum expectation value are checked in both the gauge theory side and the gravity side. We also calculate the correlation functions with the chiral primary operators in both sides and find good agreement.
Next we consider more general dilatation invariant BPS surface operators. The gravity duals of those operators are proposed.
\end{abstract}

\vfill
\vskip 0.5 truecm

\setcounter{footnote}{0}
\renewcommand{\thefootnote}{\arabic{footnote}}

\newpage

\tableofcontents

\section{Introduction} 
A surface operator in a gauge theory is an operator supported on two-dimensional surface $ \Sigma$. In  $\N=4$ super Yang-Mills theory, a disorder type surface operator is introduced to give a gauge theory description to ramifications in the context of the geometric Langlands program in number theory\cite{Gukov:2006jk}. The surface operator is characterized by the boundary condition on the fields in the path integral near a codimension two singularity. As other local or nonlocal operators, surface operators are useful to understand the AdS/CFT correspondence\cite{Maldacena:1997re,Gubser:1998bc,Witten:1998qj}. 
  
  The surface operator given in \cite{Gukov:2006jk} is half BPS and the singularity is in the form of a simple pole. 
The gravity dual of the surface operator can be studied. In \cite{Gukov:2006jk}, the gravity dual of it has been proposed as a probe D3-brane wrapping $AdS_3 \times S^1$ in $ AdS_5 \times S^5$, which can be supersymmetric \cite{Skenderis:2002vf, Constable:2002xt}. The corresponding type IIB super-gravity solution, named as bubbling geometry, has been analyzed in \cite{Lin:2004nb,Lin:2005nh,Gomis:2007fi}.   Some observables related to this surface operator are calculated in the various pictures\cite{Drukker:2008wr}.
The half BPS surface operator is generalized to the case that the singularity is a higher order pole \cite{Witten:2007td} and a simple pole up to a logarithm \cite{Gukov:2008sn}. Other kinds of surface operators are also investigated in \cite{Harvey:2007ab,Buchbinder:2007ar,Harvey:2008zz}.

One of the most interesting aspects of the surface operator of \cite{Gukov:2006jk} is the fact that some of the physical quantities may be compared between the gauge theory side and the gravity theory side. Usually it is not easy to compare those quantities because the classical gravity calculation is only valid in large 't Hooft coupling $\lambda$ regime, while the perturbative gauge theory calculation is only valid when $\lambda$ is small. However the surface operator has a parameter $\beta$, and the physical quantities can sometimes be expressed as the power series in $\lambda/\beta^2$ on the gravity theory side, which for large $\beta$ mimics the perturbative small $\lambda$ expansion\footnote{This does not mean that they must agree with each other unless the AdS/CFT correspondence is wrong. There could be ``discrepancy'' since the order of the limit is different in each side. Actually similar ``discrepancy'' happens in the anomalous dimension of large R-charge local operators in the context of the integrability in the AdS/CFT correspondence\cite{Serban:2004jf}.}. This situation is similar to what happens in the plane wave limit in \cite{Berenstein:2002jq}; the R-charge $J$ plays a similar role to $\beta$ in this case.

There are many possible ways of constructing more general surface operators preserving fewer supercharges. We restrict our attention to operators which are scale-invariant, so the locus of the singularity is a collection of planes intersecting at a single point. As we shall see, the allowed singularities may have branches.  An objection can be that the configuration is not well-defined since the boundary condition is not single valued. We show that via an appropriate gauge transformation, the possible monodromy can be canceled. In the gauge theory, a surface operator of this kind can be constructed by using homogeneous algebraic equations. The most general case becomes $1/16$ BPS. We propose that the gravity dual of it is a D3-brane wrapping a holomorphic surface $ \Sigma_4$ in $AdS_5 \times S^5$, where $ \Sigma_4$ is defined by the same homogeneous algebraic equations. We take a 1/4 BPS example to investigate the preserved symmetries, the vacuum expectation value, and the correlation function with a chiral primary operator. 
    
 The proposed D3-brane dual to the 1/4 BPS surface operator is shown to preserve 1/4 of the super symmetries in type IIB.  In the semi-classical limit, we show that the vacuum expectation value of the surface operator is 1 in both sides. In the gauge theory, we proceed to calculate the correlation function between the surface operator and local operators. In the gravity, we take the supergravity limit and present the result for all orders in $ \lambda/\beta^2$ as an integral form. Analytic results of the integration are given the leading and the next-to-leading order in $ \lambda/\beta^2$.  The leading order result coincides with that of the gauge theory.

The organization of this paper is as follows. 
In section \ref{sec:gauge theory}, we construct a specific example of the surface operator in the gauge theory.  We show that the surface operator is quarter BPS. In the semi-classical limit, we study the vacuum expectation value of the operator and correlation functions with CPO's. 
 In section \ref{sec:gravity theory}, we propose the gravity dual of this operator, and then check the preserved super symmetries by kappa symmetry projection in the embedding space.  For the D3-brane solution, we evaluate the vacuum expectation value and correlation functions with CPO's.  
 In section \ref{sec:general}, we consider a generalization of our example. Section \ref{sec:discussion} is devoted to discussions.

\section{An example of $1/4$ BPS surface operator in the gauge theory} 
\label{sec:gauge theory}

\subsection{Definition of the surface operator by a classical solution}
\label{sec:1/4def}

In this section, we will consider a surface operator $\cO_{ \Sigma}$ of  $SU(N)$ $\N=4$ super Yang-Mills theory on $\Rb^4$  with coordinates $ (x^0, x^1, x^2, x^3)$, or on $ \Cb^2$ with coordinates $ (z^1, z^2) $. Our conventions for the gauge theory are collected in appendix \ref{gauge}. 

As in \cite{Gukov:2006jk}, we characterize a surface operator by the boundary condition of bosonic fields near codimension 2 singularities. Semi-classically the surface operator is given by a classical solution with these boundary conditions and its quantum fluctuations. Likewise, any classical solution with a codimension 2 singularity corresponds to a surface operator, defined by the boundary conditions near the singularity of the classical solution.

In this paper we focus on the classical solutions in which the gauge fields are flat; in a suitable gauge choice we can set $A_{ \mu}=0$ at least locally. The non-trivial field excitations in the classical solution are the scalar fields. In the 1/2 BPS case \cite{Gukov:2006jk,Gomis:2007fi} the solution with simple pole
\begin{align}
 \Phi\sim \frac{1}{(z^1)},
\end{align}
is considered. The higher order poles can also be considered \cite{Witten:2007td}. Then what happens when the singularity has branches? That is what we address in this paper.

In order to explain our basic idea and make things explicit, we focus in this section and next on the classical solution of the type
\begin{align}
 \Phi\sim \frac{1}{\sqrt{z^1z^2}}. \label{sqrt}
\end{align}
 We will consider more generic classical solutions in section \ref{sec:general}.
In particular, other examples of the 1/4 BPS surface operators are found in section \ref{sec:general 1/4}.
In eq.\eqref{sqrt} we include both $z^1$ and $z^2$ in order to preserve the dilatation symmetry. This dilatation symmetry is useful for Wick rotation as we will explain later. 
This configuration is singular along the two planes with $z^1=0$ as well as $z^2=0$. Hence when we consider the surface operator by the path integral, we impose the boundary condition at both $z^1=0$ and $z^2=0$.

\begin{figure}
 \begin{center}
  \includegraphics[width=6cm]{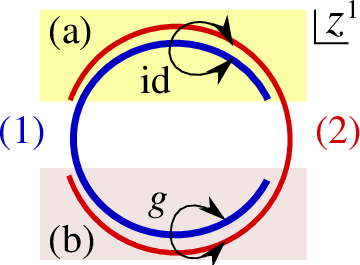}
 \end{center}
\caption{The patches of the coordinates.  We introduce two patches (1) $0<\phi_1<2\pi$, and (2) $-\pi<\phi_1<\pi$. They are connected by the gauge transformation with the identity matrix in (a) $0<\phi_1<\pi$ and the matrix $g$ in (b) $-\pi<\phi_1<0$.}
\label{fig:patches}
\end{figure}

The behavior of the scalar field \eqref{sqrt} does not look like a consistent configuration because it is not single valued\footnote{Similar double valued configuration also appears in the conformal vortex loop operator in 3-dimensional super-conformal Chern-Simons theory\cite{Drukker:2008jm}.}.  However we can make it a consistent configuration by introducing the gauge field holonomy as follows.
We consider the scalar $\Phi$, which is an $N\times N$ matrix, and the gauge fields
\begin{align}
 \Phi =  \diag \left(    \frac{  \beta  }{ \sqrt{ z^1 z^2 } } , 
   - \frac{   \beta   }{ \sqrt{z^1 z^2}} , 0, \cdots, 0  \right) ,
\qquad
A_{\mu}=0,
 \label{quarter:op}
\end{align}
where $\beta$ is a real positive parameter.
These fields are not single valued. For example, the value of the scalar field has $\Zb_2$ monodromy when $z^1$ goes to $z^1e^{2\pi i}$. This monodromy can be cancelled by the gauge holonomy. Let us introduce two patches of coordinates ($z^1=r_1e^{i\phi_1}$) in order to explain this holonomy (see figure \ref{fig:patches}).
\begin{enumerate}
 \item[(1)] $0<\phi_1<2\pi$ (branch cut at $\phi_1=\pi$).
 \item[(2)] $-\pi<\phi_1<\pi$ (branch cut at $\phi_1=0$).
\end{enumerate}
The intersection of these two patches is two disconnected regions: (a) $0<\phi_1<\pi$ and (b) $-\pi<\phi_1<0$. The gauge transformation between these two patches is chosen as follows. In the region (a), (1) and (2) are trivially identified, namely $\Phi^{(1)}=\Phi^{(2)},\ A_{\mu}^{(1)}=A_{\mu}^{(2)}$. On the other hand, in the region (b) they are related by the gauge transformation with the constant parameter $g\in$SU($N$) defined as
\begin{equation}
 g = \begin{pmatrix}  i \sigma_1 &  0  \\
0 & I_{N-2}  \end{pmatrix}.  \label{gauge_transform}
\end{equation} 
The fields are transformed under the transformation as,   
\begin{equation}
  g \Phi^{(1)} g^{-1} = - \Phi^{(1)}=\Phi^{(2)} ,   \qquad
g ( A^{(1)}_{\mu} - i \del_{\mu} ) g^{-1} = A^{(2)}_{\mu} . \label{eq:2}
\end{equation}

The eq.(\ref{eq:2}) are consistent with $A^{(1)}_{\mu}=A^{(2)}_{\mu}=0$ because $g$ is a constant. Moreover the monodromy due to the square root branch can be canceled by this gauge transformation. 

Note that the configuration \eqref{quarter:op} satisfies the equation of motion.

We can introduce further gauge holonomy which commute with $\Phi$ and $g$. The field configuration \eqref{quarter:op} and the holonomy \eqref{gauge_transform} break the gauge symmetry SU$(N)$ to U$(1)\times$SU$(N-2)$. Therefore we can introduce extra holonomy included in this U$(1)$. In other words, we can change the gauge transformation \eqref{gauge_transform} to $g_{\alpha_1}$ which includes parameter $\alpha_1$ as
\begin{equation}
 g_{\alpha_1} = \begin{pmatrix}  ie^{i\alpha_1} \sigma_1 &  0  \\
0 & e^{-2i\alpha_1/(N-2)}I_{N-2}  \end{pmatrix}. 
\end{equation} 
This parameter $\alpha_1$ is an analogue of the parameter $\alpha$ in \cite{Gukov:2006jk}.  There is also a monodromy around $z^2=0$, and the holonomy $g_{\alpha_2}$ is introduced to cancel this monodromy.  These two parameters $(\alpha_1,\alpha_2)$ fix the holonomy globally\footnote{This is so simple in this special example because the monodromy group is abelian. It is also worth to note that the fundamental group of the space, obtained by removing two planes $z^1=0$ and $z^2=0$ from $\Cb^2$, is an abelian group $\Zb\times \Zb$. In the more general case, the problem seems to be more involved.}.

We can also introduce the two dimensional theta angle as the similar way as in \cite{Gukov:2006jk} by inserting the operator
\begin{equation}
 \exp\left[i\eta_1\int_{z^1=0}F|_{{\rm U}(1)}\right],
\end{equation}
where $F|_{{\rm U}(1)}$ are the unbroken U$(1)$ part of the field strength. There are also similar parameter $\eta_2$ at $z^2=0$.

We can also introduce the parameter $\gamma$ as in \cite{Gukov:2006jk}, by making $\beta$ complex valued. However, to avoid undue complications, we choose $\beta$ as a real positive number. The phase can be restored to nontrivial value at any stage of our discussion.  

Instead of introducing two patches and a gauge transformation between them, we can choose the gauge field $ A$ to be nontrivial, $ A = - \frac{i}{ 2 \pi} \sum_{i=1,2}  \ln g_{ \alpha_i} d \phi_i $. In this frame, the scalar field takes the form 
\be
\Phi= \begin{pmatrix} (z_1z_2)^{-1/2}\sigma_3 e^{i\sigma_1(\phi_1+\phi_2)/2} & 0  \\  0 &  0_{N-2} \end{pmatrix} , \nn \ee
 and there is no monodromy. However we will not use this frame in the rest of this paper since this frame is not convenient to see the supersymmetry.

Let us consider how to do the path-integral around this multi-valued configuration. In string theory, a similar situation occurs when one consider the twisted sector of strings in the presence of orbifolds. The boundary condition of the fluctuation $ \delta \Phi $ can be chosen by the following 
\be  \left( \Phi + \delta \Phi  \right) ( \phi_i + 2 \pi ) =  g_{\alpha_i} \left( \Phi + \delta \Phi  \right)  ( \phi_i  )   g^{-1}_{ \alpha_i} , \quad  (\mbox{not summed over } i ) , \label{bdry} \ee
where $ \Phi$ is  given in \eqref{quarter:op}, a solution of the equation of motion.
Expand $ \delta \Phi $ as
\begin{align}
 \delta \Phi & = \sum_{A}T_{A}\delta\Phi^{A},
 \nn 
 \end{align}
where $T_A$ are the basis of the $N\times N$ matrix, which diagonalize the adjoint $g_{\alpha_i}$ action, namely $g_{\alpha_i}T^{A}g_{\alpha_i}^{-1}=\exp(2\pi i\xi_i(A))T^{A}$ with some numbers $\xi_i(A)$. $ \delta \Phi^{A}$ can be expanded to the Fourier series as $\delta\Phi^{A}=\sum_{n_1,n_2\in \Zb}\exp[i\sum_i(n_i+\xi_i(A))\phi_i]h^{A}_{n_1,n_2}$ and the measure of the path-integral can be written as $ D\delta\Phi=\prod_{A,n_1,n_2}Dh^{A}_{n_1,n_2}$. The other fields are also treated as the same way.

The configuration \eqref{quarter:op} preserves the dilatation symmetry. This dilatation symmetry acts on the scalar field with a real positive parameter $\alpha$ as
\begin{align}
(z_1,z_2)\to (z_1',z_2')=(\alpha z_1,\alpha z_2),\qquad
 \Phi(z^1,z^2)\to \Phi'(z_1',z_2')=\alpha^{-1} \Phi(z^1, z^2).
\label{dilatation}
\end{align}
Under the dilatation symmetry, the configuration \eqref{quarter:op} is invariant, i.e.\ $\Phi'(z_1,z_2)=\Phi(z_1,z_2)$. The dilatation symmetry is useful to write an analogous configuration in Lorentzian signature. By a Weyl transformation, flat Euclidean space $\Rb^4$ can be mapped into $\Rb\times S^3$. Dilatation transformations in $\Rb^4$ are mapped to time translations in $\Rb\times S^3$, so in this frame, the configuration \eqref{quarter:op} is time independent. Thus we can safely perform the Wick rotation of $\Rb$ direction and get a static configuration of $\N=4$ SYM in $\Rb_{\rm time}\times S^3$. This fact is useful to find a gravity counterpart in Lorentzian global $AdS_5\times S^5$.

\subsection{Supersymmetry in the gauge theory}
\label{sec:gauge SUSY}
The surface operator defined in (\ref{quarter:op}) preserves 1/4 of the supersymmetry and the super-conformal symmetry. To see this, we need to consider the variation of the fermion $ \psi$ of $\N=4$ SYM,  given in (\ref{var:fermion}). 
For the background field as in (\ref{quarter:op}), the variation can be written conveniently if we use the complex coordinates: 
\begin{equation} \delta \psi = \half ( \delb_{\bar i}  \bPhi \gamma^{ \bar{i} } \gamma^3 + \del_{i}  \Phi \gamma^{i} \gamma^{ \bar{3} }  ) \epsilon (z) - (  \bPhi \gamma^3  + \Phi \gamma^{\bar{3}}  ) \epsilon_1, \label{c:var} 
\end{equation} 
where $ \bPhi $ is the hermitian conjugation of $\Phi$.  $\epsilon(z)$ combines the parameters of super-Poincar\'e transformations $\epsilon_0$ and of super-conformal transformations $\epsilon_1$, each of which is a 16 component spinor, as follows
\begin{equation}
\epsilon(z)  
= \epsilon_0 + \eta_{i \bar{j} } ( z^i \gamma^{ \bar{j} } + \gamma^{i} \bar{z}^{ \bar{j} } ) \epsilon_1.  \label{epsilon(z)} 
\end{equation}

The variation of the fermion (\ref{c:var}) vanishes, if we impose the following condition on $\epsilon_i$, 
\begin{equation} 
\gamma^{\bar{1}}\gamma^{3}\epsilon_i= \gamma^{1 } \gamma^{ \bar{3} } \epsilon_i 
=\gamma^{\bar{2}}\gamma^{3}\epsilon_i= \gamma^{2} \gamma^{ \bar{3} } \epsilon_i 
= 0, \quad (i=0,1), \label{cond} \end{equation} 
or equivalently
\begin{equation}
 (1+\Gamma^{0145})\epsilon_i=(1+\Gamma^{2345})\epsilon_i=0.
\end{equation}
In the derivation of this condition, we use the relation
\begin{equation}
  z^i \partial_{ i} \Phi = - \Phi , \nn
\end{equation} 
which holds because  $\Phi$ is a degree $(-1)$ homogeneous function of $z^1, z^2$. 

The bosonic unbroken symmetries for the quarter BPS surface operator (\ref{quarter:op}) are as follows. The surface operator in consideration (\ref{quarter:op}) preserves the dilatation symmetry $\Rb_+$ of the 4-dimensional Euclidean conformal symmetry SO$(1,5)$. It is also invariant under an $SO(4)$ subgroup of the R-symmetry SO$(6)$.  It breaks spacetime rotational symmetry, while it preserves SO$(2)_{a}$, and SO$(2)_{b}$ symmetry which are the combinations of the spacetime rotation and R-rotation; SO$(2)_{a}$ is the diagonal part of SO$(2)_R \times $SO$(2)_{01} \times $SO$(2)_{23}$, while SO$(2)_b$ is the difference of SO$(2)_{01}$ and SO$(2)_{23}$.  The super charge in the $ ({\bf 4},{\bf 4})$ representation of SO$(4,2)\times$SO$(6)$ is reduced to $( { \bf 2}, 1)$ and $(1,{\bf 2})$ of the SU$(2) \times $SU$(2) \sim $SO$(4)$.

\subsection{Vacuum expectation value} 
In this section, we will consider the expectation value of the quarter BPS surface operator, $ \cO_{ \Sigma}$, defined in (\ref{quarter:op}). We expect this expectation value to be $1$ due to the supersymmetry. The expectation value is defined as the path integral with the boundary condition at the singularity. This path integral is approximated by the classical SYM action:
\begin{equation} 
 \langle \cO_{ \Sigma } \rangle \equiv \int_{\text{boundary condition}} [DAD\psi D\phi]e^{-S}\;
\cong \exp ( - S ) \vert_{ \Sigma }. \nn 
\end{equation} 
The relevant part of $\N=4$ SYM action in (\ref{action}) is 
\begin{equation} 
 S  =  \frac{1}{ 4 g^2} \int d^2 z^1 d^2 z^2 \eta^{i \bar{j} } \tr \left(  \del_{i} \Phi  \delb_{ \bar{j}} \bar{ \Phi}  + \delb_{ \bar{j} }  \Phi \del_i  \bar{ \Phi}  \right). \nn 
\end{equation} 
In the presence of the surface operator as in (\ref{quarter:op}), it leads to the following: 
\begin{equation} 
 S \vert_{ \Sigma}  =   \frac{ \beta^2 }{ 4 g^2} \int d^2 z^1 d^2 z^2    \left( \vert z^1 \vert^2 + \vert z^2 \vert^2 \right) \vert z^1 z^2 \vert^{-3} 	 . \nn 
\end{equation} 
Let us use the polar coordinate,  $ z^{i} = r_i e^{ i \phi_i} $ and  regulate $ r_i \in (r_0, \infty)$ for $i = 1,2$ . Then 
\begin{equation} 
S \vert_{ \Sigma}
=  \frac{\beta^2}{g^2} \left[
         \int dr_1d\phi_1d\phi_2 \frac{ 1 }{r_0} 
         +\int dr_2d\phi_1d\phi_2 \frac{ 1 }{r_0} 
\right].  \label{on-shell-bulk-S}
\end{equation}  
Our conventions for the measure are given in (\ref{measure}). 
As in \cite{Drukker:2008wr}, we add a boundary term to impose the appropriate boundary condition.  Without additional boundary terms, the variation of the action gives
\begin{multline} 
\delta S \supset -\frac{1}{2 g^2}    \tr  \Bigg[ 
\int_{r_1=r_0} d^2 z^2 d \phi_1 r_1 \left( \delta \Phi  \frac{\del}{\del r_1} \bPhi 
                          + \delta \bPhi \frac{\del}{\del r_1} \Phi \right) \\
+ \int_{r_2=r_0} d^2 z^1 d \phi_2 r_2 \left( \delta \Phi \frac{\del}{\del r_2} \bPhi
                     + \delta \bPhi \frac{\del}{\del r_2}  \Phi \right) \Bigg]. \label{surface_term}
\end{multline} 
This surface term would impose the boundary conditions $r_1 \del \Phi/ \del r_1=0$ at $r_1=r_0$ and $r_2 \del \Phi/ \del r_2=0$ at $r_2=r_0$, which are not satisfied by the solution \eqref{quarter:op}. In order to get rid of this additional condition, we add the following boundary term to the action. 
\begin{equation} 
 S_b = -\frac{1}{ 4 g^2} \tr  \left[  \int_{r_1=r_0} d^2 z^2 d \phi_1 \left( \Phi \bPhi \right)  + \int_{r_2=r_0} d^2 z^1 d \phi_2  \left(  \Phi \bPhi  \right) \right].
\nn
\end{equation} 
Adding this boundary term makes the boundary conditions $2 r_1 \del \Phi/ \del r_1+\Phi=0$ at $r_1=r_0$ and $2r_2 \del \Phi/ \del r_2+\Phi=0$ at $r_2=r_0$, which are actually satisfied by the solution \eqref{quarter:op}.

The total action is summed up to be zero, $( S + S_b) \vert_{ \Sigma} = 0$, thus
\begin{equation} 
 \langle \cO_{ \Sigma} \rangle = 1. \label{vev}
\end{equation} 

 As considered in \cite{Berenstein:1998ij,Graham:1999pm,Henningson:1999xi,Gustavsson:2003hn,Gustavsson:2004gj}, the surface operator may have conformal anomalies since the surface operator is defined on an even dimensional submanifold. To compute the anomaly, we need to evaluate the action of a surface operator defined on $ \Sigma$ with non-trivial curvature or Weyl tensor. This will be an interesting future work.

\subsection{Correlation functions with chiral primary operators}  
\label{sec:cpo gauge}

The correlation function of a local operator $\cO(\zeta)$ inserted at the point $z^m=\zeta^m$ and the surface operator $ \cO_{ \Sigma}$ in the semi-classical limit is given by the classical value of the operator in the classical solution. 
\begin{equation} 
 \frac{ \langle \cO_{ \Sigma} \cdot \cO(\zeta) \rangle }{ \langle \cO_{ \Sigma} \rangle } =\frac{1}{ \langle \cO_{ \Sigma} \rangle }\int_{\text{boundary condition}} [DAD\psi D\phi]\;\cO(\zeta)\;e^{-S}\;\cong \cO \vert_{ \Sigma } (\zeta). \label{correlation}  
\end{equation} 
The correlation function of the surface operator \eqref{quarter:op} and a chiral primary operators is non-trivial when the CPO is invariant under $SO(4)$ subgroup of $SO(6)$ R-symmetry. We use the  notation of the $SO(4)$ invariant CPO  as given in \cite{Skenderis:2007yb,Drukker:2008wr}
\begin{equation}
  \cO_{ \Delta , k } = \frac{ ( 8 \pi^2)^{ \Delta /2}}{ \lambda^{ \Delta /2} \sqrt{ \Delta} } C^{ i_1, \cdots, i_{ \Delta} }_{ \Delta, k } \tr ( \phi_{ i_1} \cdots \phi_{ i_{ \Delta}} ). 
\end{equation}
$ \Delta $ is the conformal dimension. $k$ is the charge under $SO(2)$ of $SO(2) \times SO(4)$ subgroup of R-symmetry,  $k = - \Delta, - \Delta+2, \cdots, \Delta $. We give the definition of $C^{ i_1, \cdots , i_{ \Delta} }$ in (\ref{def:C}) and review the relevant aspects of the spherical harmonics in appendix \ref{spherical}. The relevant part  for the correlation function with the surface operator in (\ref{quarter:op}), 
\begin{equation} 
\cO_{ \Delta , k } = \frac{ ( 8 \pi^2)^{ \Delta /2 } }{ \lambda^{ \Delta /2} \sqrt{ \Delta } } C_{ \Delta ,k }  \tr [ \Phi^{ ( \Delta + k )/2} \bPhi^{ ( \Delta - k )/2} + \Phi_2(\dots)+\Phi_3(\dots)]_{\text{sym}} , 
\label{SO(4)inv-cpo}
\end{equation} 
where $C_{ \Delta , k}$ is given in (\ref{def:cdk}), and $[\cdot]_{\text{sym}}$ means all the products inside are totally symmetrized. Using $\Phi_2=\Phi_3=0$ the correlation function evaluated by (\ref{correlation}) is 
\begin{equation} 
\frac{\langle \cO_{ \Delta , k } (\zeta)\cdot \cO_{ \Sigma} \rangle}{ \langle \cO_{ \Sigma} \rangle } =   \frac{ ( 8 \pi^2)^{ \Delta /2 } }{ \lambda^{ \Delta /2} \sqrt{ \Delta } } C_{ \Delta, k }  \frac{ \beta^{ \Delta} }{ \vert \zeta^1 \zeta^2 \vert^{ ( \Delta -k) / 2 } ( \zeta^1 \zeta^2)^{ k /2 } } \left( 1 + ( -1)^{ \Delta } \right) . 
\label{cpo:gauge} 
\end{equation} 
Note that the correlation function vanishes for odd $ \Delta$, since the two diagonal components inside the trace have the opposite sign.

The correlation function of the $ \cO_{ \Sigma}$ in (\ref{quarter:op}) with a Wilson line or the stress-energy tensor can be interesting physical quantities to investigate, as for the half BPS surface operator in \cite{Drukker:2008wr}. However we will not pursue the issue here. 

\section{Gravity dual of the $1/4$ BPS surface operator}
\label{sec:gravity theory}

\subsection{Probe D3-brane as the gravity dual of the surface operator}
 
 Let us now consider the holographic dual of the surface operator defined in (\ref{quarter:op}).  The following complex coordinate system for  $AdS_5 \times S^5 $ is convenient for this purpose: 
\begin{equation}
 ds^2 = \frac{1}{  \sum_a \vert \omega^a \vert^2 } \left( \sum_{a=1}^3  \vert d \omega^a \vert^2 +   (  \sum_a \vert \omega^a \vert^2  )^2    \sum_{m=1}^2 \vert dz^m \vert^2 \right). \label{metric} 
\end{equation}
We can relate it to the global coordinates of $AdS_5\times S^5$ as follows: 
\begin{eqnarray}
 \omega^a &=& e^{ - \tau} \csc \rho u_a \exp(i\theta_a), \quad \sum_{a=1}^3 u_a^2 = 1, \nn \\
  z^m &=& e^{ \tau} \cos \rho r_m \exp(i\phi_m), \quad \sum_{ m=1}^2 r_m^2 = 1, \nn 
\end{eqnarray} 
Then the metric \eqref{metric} becomes
\begin{equation}
 ds^2  = \frac{1}{ \sin^2 \rho} \left( d \tau^2 + d \rho^2 + \cos^2 \rho d \Omega_3^2 \right)  + d\Omega_5^2,  \nn 
\end{equation} 
where $d\Omega_3^2$ and $d\Omega_5^2$ are the metrics of the unit $S^3$ and the unit $S^5$ expressed as
\begin{equation}
 d\Omega_3^2=\sum_{m=1,2}(dr_m^2+r_m^2 d\phi_m^2),\qquad
 d\Omega_5^2=\sum_{a=1,2,3}(du_a^2+u_a^2 d\theta_a^2).
\end{equation}
To be supersymmetric, we have the following 5 form field strength in the background: 
\begin{equation}
 F_5 = 4 ( vol ( AdS_5) + vol ( S^5 ) ). \label{flux} \end{equation} 
In this paper we choose the unit of length such that the radius of $AdS_5$ is $1$, namely $4\pi g_s N \alpha'^2=1$. In these units $\alpha'=1/\sqrt{\lambda}=1/\sqrt{4\pi g_s N}$, and the D3-brane tension $\TD$ is expressed as
\begin{equation}
  \TD=\frac{1}{(2\pi)^3 g_s \alpha'^2}=\frac{N}{2\pi^2}.
\label{TD3}
\end{equation}

We propose that a probe D3-brane, wrapping a 4 dimensional subspace $ \Sigma_4$ defined by the following holomorphic equations , 
\begin{equation} 
 z^1 z^2 (  \omega^1 )^2 - \kappa^2 =0, \quad  \omega^2 = \omega^3  = 0 ,
\label{wrapping} 
\end{equation} 
is dual to the surface operator in (\ref{quarter:op}). Here $ \kappa$ is a parameter related to $\beta$.
The precise relation between $\beta$ and $\kappa$ are determined later in eq.(\ref{identification}), according to the correlation function with the chiral primary operators.

The parameters $(\alpha_1,\alpha_2,\eta_1,\eta_2)$ are mapped to the gauge field $A$ on the D3-brane and its magnetic dual $\tilde{A}$ as
\begin{align}
 A=\alpha_1 d\phi_1+\alpha_2 d\phi_2,\qquad
 \tilde{A}=\eta_1 d\phi_1+\eta_2 d\phi_2.
\end{align}

Note that the configuration \eqref{wrapping} is $\tau$ independent, i.e.\ static. Thus we can Wick rotate the configuration easily by replacing $\tau=it$ and consider the same static configuration in the Lorentzian signature. It is convenient to consider the Lorentzian signature especially when checking the supersymmetry, which we do in the next subsection.
\subsection{Supersymmetry of the probe brane} 
\label{sec:gravity SUSY}
To check the preserved supersymmetry of the probe D-brane,  we can use the kappa symmetry projection \cite{Cederwall:1996pv,Aganagic:1996pe,Cederwall:1996ri,Bergshoeff:1996tu,Aganagic:1996nn,Bergshoeff:1997kr,Skenderis:2002vf}.   To this end, it is advantageous to use a 12 dimensional embedding space $ \Cb^{ 1, 2} \times \Cb^3 $ with coordinates $ ( Z^0, Z^1, Z^2, W^1, W^2, W^3) $, as in \cite{Mikhailov:2000ya, Kim:2006he}. The reason is that the Killing spinor of the embedding space is constant Dirac spinor with 64 complex components, while the Killing spinor of the physical space depends on the space-time.  We can get the 10 dimensional physical space  $AdS_5 \times S^5$ using the constraints: 
\begin{equation}    - \vert Z^0 \vert^2 + \vert Z^1 \vert^2 + \vert Z^2 \vert^2 = - 1, \quad
 \sum_{ A= 1,2,3}  \vert W^A \vert^2   = 1 .  
\label{physical}
\end{equation} 
Reduction of the constant Killing spinor in the embedding space to the one in the physical space is done by the following conditions, 
\begin{equation} 
 8 \gamma_{ 0  \bar{0}   } \gamma_{ 1  \bar{1}  }  \gamma_{ 2 \bar{2}   } \epsilon = \epsilon,  \quad
  8 \gamma_{ 3 \bar{3}    } \gamma_{ 4 \bar{4}  }  \gamma_{ 5 \bar{5}   } \epsilon = \epsilon,
\label{weyl}  
\end{equation} 
where $ \gamma^A$ are defined in (\ref{cplx_gamma}). 

We review the relevant aspects of the kappa symmetry projection and  the conventions for the embedding space  in appendix \ref{kappa} and appendix \ref{12d_convention}. General discussions about super-symmetric branes in $AdS_5 \times S^5$ can be found in \cite{Skenderis:2002vf}.

We can use an embedding map as follows: 
\begin{equation} 
\begin{split}
Z^0 &= \csc \rho e^{ i t },  \\ 
Z^m &= \cot \rho r_m e^{ i \phi_m} =   \left( e^{ it} \sin \rho  \right)^{-1} z^m ,  \quad ( m = 1,2),
\\
W^a &= u_a e^{ i \theta_a} =  e^{  it } \sin \rho  \omega^a, \quad ( a = 1,2,3), 
\end{split}
\label{embedding_map}
\end{equation} 
The D3-brane worldvolume $ \Sigma_4$ defined by (\ref{wrapping}) is an intersection of the physical space and a six dimensional holomorphic space $ \Sigma_6$ in the embedding space defined by: 
\begin{equation} 
Z^1 Z^2 ( W^1)^2 - \kappa^2  = 0, \quad  W^2 = W^3 = 0  .
\label{12d_ftns} 
\end{equation} 
Note that we can let eq.(\ref{12d_ftns}) be the same form of eq.(\ref{wrapping}) due to our embedding map (\ref{embedding_map}).  

 Two vectors $ E_{r_1}, E_{r_2} $ in $ T \Sigma_6 \backslash T  \Sigma_4$ are projections of the two normal vectors of $ AdS_5 \times S^5 $ in the embedding space on the $ T \Sigma_6$. One of these vectors is time-like and the other is space-like. We will use $ r_1 $ as a time-like direction. 
 
Since  $ T \Sigma_6$ is closed under a complex structure $I$, given as 
$ I \cdot { \frac{ \partial}{ \partial  Z^A } } =  i  \frac{ \partial}{ \partial Z^A}$ and $I \cdot \frac{ \partial}{ \partial  \bar{Z}^{ \bar{A}} } = -  i  \frac{ \partial}{ \partial  \bar{Z}^{ \bar{A}} } $
, $I \cdot E_{r_i}  $ are in $ T \Sigma_4$ .  We call these vectors $ E_0, E_1 $. We can linearly combine $E_0, E_1$ to form  null vectors $E_{ \pm}$ .  There leave two linearly independent vectors in $ T \Sigma_4$, which are orthogonal to $ E_{\pm} $ and closed under $I$.  The holomorphic/anti-holomorphic part of these vectors are defined to be  $ E_{ z}, E_{ \bar{z} }$. From the construction, the followings hold: 
\begin{equation}  E_z^{ \bar{A}} = 0,  \quad  \eta_{A \bar{B} } E_z^{A} E_{ \pm}^{ \bar{B} } = 0 , \quad \eta_{A \bar{B} } E_{+}^A E_{+}^{ \bar{B} } = \eta_{A \bar{B} } E_{-}^{A} E_{-}^{ \bar{B} } = 0 . 
\label{tangent:1}  
\end{equation} 
We normalize the vectors as: 
\begin{equation} 
 \eta_{A  \bar{B} } E_z^{ A} E_{ \bar{z}}^{ \bar{B} }  =  \eta_{ z \bar{z} } = \half   , \quad
 \eta_{ A  \bar{B} } E_{+}^A E_{-}^{ \bar{B} }  = \eta_{ A \bar{B} } E_{-}^A E_{+}^{  \bar{B} } =  \frac{1}{4}   . 
 \label{tangent:2}
 \end{equation} 
Using eqs.\eqref{tangent:1},\eqref{tangent:2}, the projection operator (\ref{projection}) along the D3-brane can be written in the following form: 
\begin{equation} \Gamma = 4 i \left( ( E_+^B \gamma_B + E_+^{ \bar{B} } \gamma_{ \bar{B} } ) ( E_{-}^{C} \gamma_C + E_{-}^{ \bar{C} } \gamma_{ \bar{C} } ) - \half  \right) ( E_{ \bar{z}}^{ \bar{A} } E_z^{D} \gamma_{ \bar{A} } \gamma_D  - \half ) . 
\label{hol:projection} 
\end{equation} 

As shown in \cite{Mikhailov:2000ya}, the condition for the preserved supersymmetry of a D3-brane in the embedding space becomes 
\begin{equation} \Gamma^{ r_1 r_2} \Gamma \epsilon =   i \epsilon,  \label{12d_projection} 
\end{equation} 
where $ \Gamma$ is the projection operator along the D3-brane, for our case (\ref{hol:projection}).

Let us now consider the holographic dual of the surface operator in (\ref{quarter:op}). We will use  $(Z^3,Z^4,Z^5) \equiv ( W^1, W^2, W^3)$.  For the holomorphic space $ \Sigma_6$ defined by  (\ref{12d_ftns}),   the tangent vectors are $
 v_0 \equiv Z^0  \partial_0 $, $ v_1 \equiv  \left(  Z^1 \partial_1 + Z^2 \partial_2 - Z^3 \partial_3 \right) $,   $  v_2 \equiv  \left(  Z^1 \partial_1 - Z^2 \partial_2 \right) $
and their complex conjugates. The vector  in $ T \left( \Cb^{1, 2} \times \Cb^1 \right) \backslash T \Sigma_6$ is
$ v^{ \perp} \equiv  \half \bar{Z}^3 \bar{Z}^2 \partial_1 + \half \bar{Z}^3 \bar{Z}^{ 1} \partial_2  + \bar{Z}^1 \bar{Z}^2 \partial_3 $. 
To project the normal vectors of $AdS_5 \times S^5$ on $ T \Sigma_6$, we note that on $\Sigma_6$, 
\begin{align}
Z^0 \partial_0 + Z^1 \partial_1 + Z^2 \partial_2 =& v_0 + \frac{ 4 \vert Z^1 Z^2 \vert^2 }{ \nu} v_1
+ \frac{\vert Z^1 \vert^2 - \vert Z^2 \vert^2 }{\nu} v_2 + \frac{ 4 Z^1 Z^2 Z^3 }{\nu} v^{ \perp}, \nn \\
Z^3 \partial_3 =& - \frac{ \vert Z^1 \vert^2 + \vert Z^2 \vert^2 }{ \nu} v_1 + \frac{ \vert Z^1 \vert^2 - \vert Z^2 \vert^2}{\nu} v_2 + \frac{ 4 Z^1 Z^2 Z^3}{\nu} v^{ \perp}, \nn 
\end{align}
where $\nu \equiv 4 \vert Z^1 Z^2 \vert^2 + \vert Z^1 \vert^2 + \vert Z^2 \vert^2 $. We project out $v^{ \perp}$ then  linearly recombine the result to get a convenient form of  $E_{r_1}, E_{r_2}$. We choose 
\begin{align}
 E_{r_1}  = &  \sqrt{ \frac{ \vert Z^1 \vert^2 + \vert Z^2 \vert^2 }{ \nu} } \left(  v_0 + \frac{  \vert Z^1 \vert^2 -\vert Z^2 \vert^2}{ \vert Z^1 \vert^2+\vert Z^2 \vert^2} v_2     + c.c  \right)  \nn \\
E_{r_2}  = &   \sqrt{ \frac{ \vert Z^1 \vert^2 + \vert Z^2 \vert^2}{\nu} } \left(   v_1 - \frac{  \vert Z^1 \vert^2-\vert Z^2 \vert^2}{ \vert Z^1 \vert^2+\vert Z^2 \vert^2} v_2 + c.c  \right) . 
\nn
\end{align}
We can now construct tangent vectors satisfying \eqref{tangent:1}, \eqref{tangent:2} as follows: 
\begin{align}
  E_z =& \frac{1}{ Z^0 \sqrt{ \nu} }  \left(  ( \vert Z^1 \vert^2 -\vert Z^2 \vert^2 )  v_0 +  \vert {Z}^0 \vert^2 v_2   \right)  , \quad E_{ \bar{z}} = E_z^{ \ast},  \nn \\
E_{ \pm }   =& \frac{1}{2} \left( \pm I \cdot E_{r_1} + I \cdot E_{r_2} \right) . 
\nn
\end{align}

From the construction,  $ E_{ \pm }^A \gamma_A= \frac{i}{2} ( \pm E_{r_1}^A  + E_{r_2}^A ) \gamma_A, E_{ \pm}^{ \bar{A} } \gamma_{ \bar{A} } = - \frac{i}{2} ( \pm E_{r_1}^{ \bar{A} }+ E_{r_2}^{ \bar{A} }) \gamma_{ \bar{A}} $.  Due to the orthogonality of $E_{r_1}$ and $E_{r_2}$ , the relations $ \{ E_{r_1}^A \gamma_A, E_{r_2}^{ \bar{B} } \gamma_{ \bar{B} }  \} = \{ E_{r_1}^{ \bar{A}} \gamma_{ \bar{A}} , E_{r_2}^B \gamma_B  \} = 0 $ hold. $ \Gamma^{r_1} $ is given as $ \Gamma^{r_1} = - \Gamma_{r_1} = - ( E_{r_1}^A \gamma_A + E_{r_2}^{ \bar{B} } \gamma_{ \bar{B} } ) $. Using these, the projection (\ref{12d_projection}) can be written as 
\begin{equation}
 \Gamma^{r_1 r_2} \Gamma \epsilon = -2i   \lbrack E_{r_1}^A \gamma_A , E_{r_1}^{ \bar{A}} \gamma_{ \bar{A}}  \rbrack   \lbrack  E_{r_2}^B \gamma_B , E_{r_2}^{ \bar{B} } \gamma_{ \bar{B} }  \rbrack  E_{ \bar{z} }^{ \bar{C} } E_z^D \gamma_{ \bar{C} D  } \epsilon .
\label{quarter_projection}
\end{equation} 
Let us impose the following conditions on the Killing spinor $ \epsilon $, 
\begin{equation} \gamma_{  1 \bar{1}  } \epsilon = \gamma_{  2  \bar{2 }  }  \epsilon = \gamma_{  3  \bar{3}  } \epsilon . 
\label{quarter_condition} 
\end{equation} 
It implies that $
\gamma_{1 \bar{2} } \epsilon = 
\gamma_{1 \bar{3}} \epsilon = \gamma_{2 \bar{3} } \epsilon = 0 $, since $ \gamma_{ 1 \bar{2}} \gamma_{ 1 \bar{1} }  = -  \gamma_{ 1 \bar{2} } \gamma_{ 2 \bar{2} }   =-  \half \gamma_{ 1 \bar{2} }  $  , etc. 
 It also implies that $ \gamma_{ 0 \bar{0} } \epsilon = \half \epsilon , 
 \gamma_{ \bar{0}} \epsilon = 0 ,$ when combined with the reduction of the spinor  (\ref{weyl}). Under the imposition (\ref{quarter_condition}),  the right hand side of (\ref{quarter_projection}) is reduced to $i \epsilon$, being equivalent to the condition in (\ref{12d_projection}).   It shows that the probe D-brane preserves 1/4 of the supersymmetry.

Our construction can be regarded as a special case of \cite{Kim:2006he}, in which a holomorphic space defined by three holomorphic homogeneous functions $f_k ( Z^0, Z^1, Z^2, W^1, W^2, W^3)$ for $k = 1,2,3$, when $Z^i, W^i$ are assigned weights $+1, -1$, has been considered. The probe D3-brane wrapping this holomorphic space has been shown to preserve 1/16 of supersymmetry. In our case, the $ \frac{ \partial}{ \partial t}$ vector in $AdS$ space is tangential to the D-brane, thus the transverse velocity $v$ vanishes.

\subsection{Vacuum expectation value from D3-brane action}

At the semi-classical level, we evaluate the expectation value of a surface operator by
\begin{equation} \langle \cO_{ \Sigma} \rangle = e^{ - S_{ D3} } ,
\label{vev:d3}
\end{equation} 
where $S_{ D3}$ is on-shell action of the probe D3-brane corresponding to the surface operator. This holds in the large $N$ limit ($  N \gg 1 $) and the large t'Hooft coupling limit, 
\begin{equation}  \lambda \gg 1 .  \label{limit:lambda} 
\end{equation} 

The action of the D3-brane can be expressed in terms of of $DBI$ action and Wess-Zumino term, 
\begin{equation} S_{D3} = S_{DBI} - S_{WZ}, \quad S_{ DBI} =  \TD \int d^4 \xi \sqrt{  \vert \det G_{mn}  \vert } , \quad S_{WZ} = \TD \int_{ \Sigma_4 }  C_4 . 
\nn 
\end{equation} 
$ \xi^m$ are the world-volume coordinates, and  $G_{mn}$  is the induced metric on the world-volume. $C_4$ is the R-R four form of which field strength is given is  (\ref{flux}). $\TD$ is the tension of the D3-brane (see eq.\eqref{TD3}). 

Let us consider the probe D3-brane wrapping $\Sigma_4$  defined by (\ref{wrapping}). 
We will use complex coordinates in $AdS_5 \times S^1$ with the metric \footnote{The other directions in $S^5$ are irrelevant to our discussion in this subsection} : 
\begin{equation}
 ds^2 = \vert \omega \vert^2 \sum_{m=1,2} \vert dz^m \vert^2 + \frac{ \vert d \omega \vert^2 }{ \vert \omega \vert^2 }.
\nn 
\end{equation} 
We choose the gauge of the R-R four form as
\begin{equation} 
C_4 = \frac{ \vert \omega \vert^4 }{4} dz^1 d \bar{z}^{ \bar{1} } dz^2 d \bar{z}^{ \bar{2} } . 
\nn 
\end{equation} 
We sometimes use polar coordinates $ r_i e^{ i \phi_i} = z^i$, $ \omega = y^{-1} e^{ i \theta}$. $ \theta$ is the angle of $S^1$ in $S^5$. The definition of $ y$ is to restore the Poincar\'e metric of $AdS_5$ as 
\begin{equation} 
 ds^2_{ AdS_5} = \frac{1}{y^2} \left( dy^2 + \sum_{m=1,2} (  dr^m +r_m^2 d \phi_m^2 ) \right) . 
 \nn 
 \end{equation} 
We choose $z^1, z^2$ as the world volume coordinates. 
The  transverse coordinate $ \omega$ is a holomorphic function of $z^1, z^2$.  The induced metric is 
\begin{equation} ds^2_{ind} = \frac{1}{ \vert \omega \vert^2 } \left( \sum_{m=1,2} ( \vert \omega \vert^4 + \vert \partial_m \omega \vert^2 ) \vert d z^m \vert^2  + \partial_1 \omega \delb_{ \bar{2} } \bar{ \omega} dz^1 d \bar{z}^{ \bar{2} } + \del_2 \omega \delb_{ \bar{1} } \omega dz^2 d \bar{z}^{ \bar{1} } \right).
\nn 
\end{equation} 

The evaluation of the action for this solution is given as 
\begin{equation} S_{DBI} = \frac{\TD}{4}  \int d^4 z   \left( \vert \omega \vert^4 + \vert \del_1 \omega \vert^2 + \vert \del_2 \omega \vert^2 \right)
 , \quad  S_{ WZ} = \frac{\TD}{4}  \int d^4 z  \vert \omega \vert^4  ,  \nn \end{equation} 
where the measure is defined by $ \int d^4 z \equiv \int  \vert dz^1 d \bar{z}^{ \bar{1} } dz^2 d \bar{z}^{ \bar{2} }  \vert = 4 \int d^4 x $.

 We need to add a boundary action, as in the gauge theory,  to make the variation of the action well-defined.  
The surface term which arises from the variation of the action is  
$ \delta S =   \frac{\TD}{4} \int d^4 z \sum_{m=1,2} \left( \partial_m ( \delta \omega \delb_{ \bar{m} } \bar{ \omega} ) + \delb_{ \bar{m} } \left( \delta \bar{ \omega} \del_{m} \omega \right) \right)   $. 
 Recall that the solutions of (\ref{wrapping}) are
\begin{equation}
  \omega ( z^1, z^2 ) = \pm \frac{ \kappa} { \sqrt{ z^1 z^2 } }.
\label{quarter:d3} 
\end{equation} 
For the solution, we can cancel the surface term by the following boundary action: 
\begin{equation} S_b =    \frac{\TD}{8}  \left( \int d^2 z^2 d \phi_1 dr_1  \frac{\del}{\del r_1}  \vert \omega \vert^2 + \int d^2 z^1 d \phi_2 d r_2 \frac{\del}{\del r_2}\vert \omega \vert^2   \right).   \nn \end{equation} 
 The total on-shell action of the probe D3-brane vanishes, $  ( S_{DBI} + S_b ) - S_{WZ} = 0$. The expectation value of the surface operator in the semi-classical limit is evaluated to be $1$, which coincides with the result of the gauge theory (\ref{vev}).

\subsection{Correlation functions with chiral primary operators in the gravity side} 
\label{sec:cpo gravity}
Let us now consider correlation functions of the surface operator with chiral primary operators. As in \cite{Drukker:2008wr}, we will use the GKPW prescription \cite{Gubser:1998bc,Witten:1998qj} to compute the one point function of chiral primary operators in the presence of the surface operator. We use the source 
\begin{equation}
 s = \frac{1}{4}  \int  d^4 \zeta G( y ,  z ; \zeta  )  s( \zeta  , \Omega ),  \label{source} 
\end{equation} 
where $s( \zeta, \Omega)  $ is the source at the boundary at a position of $ \zeta^m $. Let $( \zeta^1, \zeta^2) $ be $ ( d_1 e^{ i \phi_1^{ \prime}}, d_2 e^{ i   \phi_2^{ \prime} }  )$.  $ \Omega$ of interest is given as the great circle of $S^5$, $  s( \zeta, \Omega)  =  \sum_{ k } C_{ \Delta , k } e^{i  \theta}  s_0^{ \Delta, k }( \zeta )$, 
where $C_{ \Delta , k }$ is a constant defined by the $SO(4)$ invariant spherical harmonic function of $S^5$ in (\ref{def:cdk}).  $G(y,  z; \zeta )$ is the boundary-bulk propagator of a scalar field  $ s$ in $AdS_5$ with the equation of motion $ \nabla_{ \mu} \nabla^{ \mu} s = \Delta ( \Delta -4 ) s $, and is given by $ G (  y, z ; \zeta ) = \frac{ c( \Delta)  y^{ \Delta} }{ \left( y^2 + \sum_{m= 1,2} \vert z^m - \zeta^{m } \vert^2 \right)^{ \Delta}  } $. 
 $c( \Delta )$ is normalized as $ c( \Delta ) = \frac{ \Delta +1}{ 2^{ 2 - \Delta /2} N \sqrt{ \Delta } } $  \cite{Berenstein:1998ij} .  

The action of the linearized fluctuation of the D3-brane is
\begin{equation} 
\begin{split}
 \CS_{ DBI} & = \frac{\TD}{2} \int d^4 \xi \sqrt{ \det G} G^{mn} ( \partial_m X^{ \mu } \partial_n X^{ \nu} h_{ \mu \nu}^{ AdS} + \partial_m X^{ \alpha} \partial_n X^{ \beta} h_{ \alpha \beta}^S )   , \\
 \CS_{ WZ} & = \TD \int a^{ AdS} ,
 \end{split}
 \label{linear_DBI}
 \end{equation} 
where we use the notations for the fluctuation of fields, $ h_{ \mu \nu}^{ AdS}, h_{ \alpha \beta}^S, a_{ \mu \nu \rho \sigma}^{AdS}$  in  \cite{Drukker:2008wr,Kim:1985ez, Lee:1998bxa}. We use $ \mu, \nu$ indices for $AdS_5$ and $ \alpha, \beta$ for $S^5$.   We can substitute the fluctuation with the source  by using the following solution, given in  \cite{Kim:1985ez, Lee:1998bxa} : 
\begin{equation} 
\begin{split}
h_{ \mu \nu}^{ AdS} & =  - \frac{6}{5} g_{ \mu \nu}  \Delta s + \frac{4}{ \Delta +1} \nabla_{ ( \mu } \nabla_{ \nu)} s,   \\
h_{ \alpha \beta}^S &= 2 g_{ \alpha \beta} \Delta s ,    \\
a_{ \mu \nu \rho \sigma} &= -4 \sqrt{ g^{ AdS} }  \epsilon_{ \mu \nu \rho \sigma \eta } \nabla^{ \eta} s ,
\label{sugra:sol}
\end{split}
\end{equation} 
where $ \nabla_{ ( \mu } \nabla_{ \nu)} $ is the symmetric traceless part of $ \nabla_{ \mu} \nabla_{ \nu} $. The $g_{ \mu \nu}, g_{ \alpha \beta}$ are the space-time metric while $G_{mn}$ is the induced one. To get the one point function of a CPO $ \cO_{ \Delta, k }$, we take the functional derivative of $ \CS_{ D3} = \CS_{ DBI} - \CS_{ WZ}  $ with respect to $ s_0^{ \Delta, k }(\zeta) $.

Let us now consider the quarter BPS D3-brane described by $\omega(z^1, z^2) = \kappa / \sqrt{ z^1 z^2}  $ in   (\ref{quarter:d3}). The solution can be written as $ y(r_1, r_2) = \sqrt{r_1 r_2} / \kappa$, $ \theta = - \half( \phi_1 + \phi_2)$ where $ \theta$ is the argument of $ \omega$. The linearized DBI action in (\ref{linear_DBI}) can be written as
\begin{align}
 \CS_{ DBI} =& \frac{ \TD}{2} \int d^4 z  \vert \omega \vert^2  \left( ( 1+  \frac{ \vert \del_{2} \omega \vert^2 }{ \vert \omega \vert^4} ) h_{ \bar{1} 1} + ( 1 +  \frac{ \vert \del_1 \omega \vert^2}{ \vert \omega \vert^4 }  ) h_{ \bar{2} 2} +  ( \vert \del_1 \omega \vert^2 + \vert \del_2 \omega \vert^2 ) h_{ \bar{ \omega} \omega}  \right) \nonumber \\
& + \frac{\TD}{2} \int d^4 z   \left(\vert \omega \vert^2    \sum_{m=1,2}(\del_m \omega  h_{ \bar{m} \omega } + \delb_{ \bar{m} } \bar{ \omega}  h_{ \bar{ \omega} m } )
 - \frac{1}{ \vert \omega \vert^4} 
\left( \del_1 \omega \delb_{ \bar{2} } \bar{ \omega} h_{ \bar{1} 2}
 + \del_2 \omega \delb_{ \bar{1} }
 \bar{ \omega} h_{ \bar{2} 1} \right)\right),
\end{align}
where $ h_{ \bar{ \omega} \omega} = \frac{y^4}{4} ( h_{ yy}^{ AdS} + \frac{1}{y^2} h_{ \theta \theta}^S )$, etc. 
We now substitute the fluctuations with the source, using (\ref{sugra:sol}).  The Wess-Zumino term becomes $ \CS_{ WZ} = -  \TD \int d^4 z  y^{-5}  \left( \nabla^y - \frac{\partial y }{ \del r_1}  \nabla^{r_1} - \frac{ \partial y}{ \del r_2} \nabla^{r_2} \right) s $,  where $s$ is the source in (\ref{source}).  The final result is 
\begin{align} 
&\frac{ \delta  \CS }{ \delta s_0^{ \Delta, k }(\zeta) }   =  - 2 \Delta \TD \, c( \Delta ) C_{ \Delta , k } \int d^4 z \frac{  \omega^{ - \frac{ \Delta - k }{2} } (z)\bar{ \omega}^{ - \frac{ \Delta + k }{2} } ( \bar{z} )  }{ L^{ \Delta +2} }  \frac{ \vert \zeta^m \del_m \omega (z)  \vert^2 }{ \vert \omega (z) \vert^2 }  ,
\label{int:quarter} \\
& L\equiv\sum_{m=1,2}|z^m-\zeta^m|^2+|\omega|^{-2}.
\end{align} 

We will evaluate this integral in the large $\kappa$ limit
\begin{equation}
 \kappa \gg 1 \label{limit:kappa}.
\end{equation} 
In this limit,
the integration is simplified since  the integrand in (\ref{int:quarter}) becomes a delta function supported at $z^m = \zeta^m$. The limit (\ref{limit:kappa}) corresponds to the case that the slope of D3-brane approaching the boundary of $AdS_5$ becomes small, which can be seen from a rewritten solution in Poincar\'e metric, $ y (r_1, r_2)= \kappa^{-1} \sqrt{r_1r_2}  $.  In the leading order of $ \kappa$, we can replace $ \omega(z) , \del_m \omega(z) $ with $ \omega( \zeta) , \del_m \omega( \zeta)  $. 

 Here one should be careful because the coordinates $(z^1,z^2)\in \Cb^2$ do not cover the whole D3-brane worldvolume, namely the function $\omega(z)$ is double valued. Thus one needs to extend it to two $\Cb^2$'s as one usually does in the Riemann surfaces. As a result, $ z^m=\zeta^m$ is actually two points on the D3-brane worldvolume:
$y=    \kappa^{-1} \sqrt{d_1 d_2} $ with
\begin{equation}
 \theta=-\frac12(\phi'_1+\phi'_2),\qquad \text{or} \qquad 
\theta=-\frac12(\phi'_1+\phi'_2+2\pi).
\end{equation}
The result is the sum of the contributions from these two points. If we replace  $\omega(z) , \del_m \omega(z) $ with these values and use the formula \eqref{integration:1}, the integration \eqref{int:quarter} becomes 
\begin{equation}
  \frac{\langle \cO_{ \Delta , k } \cdot \cO_{ \Sigma} \rangle}{\langle \cO_{ \Sigma} \rangle}  
= - \frac{ 2^{ \Delta /2} }{ \sqrt{ \Delta} } C_{ \Delta , k } \frac{ \kappa^{ \Delta} }{(d_1 d_2)^{  \Delta /2 } } e^{ - i k ( \phi_1^{ \prime} + \phi_2^{ \prime} ) /2 } (1+(-1)^{\Delta}).  \label{cpo:gravity}
\end{equation}
This result coincide with the result in the gauge theory\eqref{cpo:gauge} if we identify the parameter $\kappa$ and $\beta$ by
\begin{equation}
 \kappa=\frac{2\pi\beta}{\sqrt{\lambda}}.\label{identification}
\end{equation}

To proceed to the next-to-leading order in $\kappa$, we expand fields near $ z= \zeta$. Let us define $v^m \equiv z^m - \zeta^m $, then $ \omega(z) = \omega( \zeta) + v^m \del_m \omega ( \zeta ) + \cdots$. Change the integration variable from $z^m$ to $v^m$. The contribution from the linear order in $v^m$ vanishes,  which can be shown  by the transformation $v^m \to -v^m$. At the quadratic order in $v^m$,  the integral of the form $ \int d^4 v \frac{ v^m v^n}{ L^{ \alpha}}$ vanishes by eq.(\ref{integration:2}). Thus, the first nontrivial term at the next order comes from the integral  of the form  $ \int d^4 v \frac{ v^m \bar{v}^{ \bar{n} } } { L^{ \alpha} } $.

 It is also useful to note that $ \zeta^m \del_m \omega ( \zeta) = - \omega ( \zeta) , \zeta^m \del_m \del_n \omega ( \zeta ) = -  2 \del_n \omega ( \zeta)  $. The integration (\ref{int:quarter}) is proportional to 
 \begin{equation} \int d^4 v \frac{1}{ \tilde{L}^{ \Delta +2} } \left( 1 + q \bar{q}  \left( \left( \frac{ \Delta^2 - k^2}{4} - \Delta +1 \right) + \frac{ ( \Delta +2)( - \Delta +1)}{ \vert \omega ( \zeta) \vert^{2}  \tilde{L} } + \frac{ ( \Delta +2)( \Delta +3) }{ \vert \omega ( \zeta) \vert^{4}  \tilde{L}^2  } \right) + \cdots  \right) ,
 \nn 
 \end{equation} 
 where $ q \equiv v^m \del_m \omega( \zeta) / \omega ( \zeta ) $ , $ \tilde{L} \equiv L ( \omega( \zeta), v ) $. We expand $L ( \omega(z), v)$ as $ L( \omega(z), v) =  \tilde{L}+ \vert \omega \vert^{-2} ( - q - \bar{q} + q \bar{q} + \cdots ) $. The evaluation of the above leads to : 
\begin{equation}
 \frac{\langle \cO_{ \Delta , k } \cdot \cO_{ \Sigma} \rangle}{\langle \cO_{ \Sigma} \rangle}  = I_0^{(1/4)} \left( 1 +  \kappa^{-2} \frac{ \Delta^2 - k^2 }{ 16 ( \Delta -1) } \left( \frac{ d_1^2 + d_2^2}{ d_1 d_2} \right) + \cdots \right), 
\label{correction} 
\end{equation} 
 where $I_0^{(1/4)}$ is the leading order result given in (\ref{cpo:gravity}).

\section{General dilatation invariant less BPS surface operator} 
\label{sec:general}

In this paper we mainly consider the 1/4 BPS surface operator of the form \eqref{quarter:op}. Actually in the gauge theory it can be generalized to those defined by a set of homogeneous algebraic equations. In this section we consider these rather general surface operators and their gravity duals.

\subsection{General less BPS surface operators in the gauge theory} \label{lessBPS}
In order to define surface operators by the boundary condition in the path integral, let us consider classical solutions with singularities as done in section \ref{sec:gauge theory} and \cite{Gukov:2006jk}.

Consider the three algebraic equations for $ \chi^a,\ a=1,2,3$, 
\begin{equation} 
\begin{split}
f_1( z^1,z^2, \chi^1,\chi^2,\chi^3) & =  0,  \\
f_2( z^1,z^2, \chi^1,\chi^2,\chi^3) & =  0,   \\
f_3( z^1,z^2, \chi^1,\chi^2,\chi^3) &= 0 , 
\end{split}
\end{equation} 
where $f_p$, $(p=1,2,3)$, are polynomial of $\chi$'s with degree $n_1$ whose coefficients are holomorphic functions of $z$'s. We also require that $f_p$ are homogeneous with weights $(+1)$ for $z$ and $(-1)$ for $\chi$. 
For fixed $z$, there are $n=n_1n_2n_3$ solutions denoted by $(\chi^1_{(i)}(z), \chi^2_{(i)}(z) , \chi^3_{(i)}(z))\quad i=1,\dots,n $. Note that $\chi^a_{(i)}(z)$ are locally holomorphic degree $(-1)$ homogeneous function of $z$. Consider a configuration of the complex scalar fields 
\begin{equation}
 \begin{split}
  \Phi_1 &= \diag ( \chi^1_{(1)}(z), \cdots, \chi^{1}_{(n)}(z), 0, \cdots, 0 ) , \\
 \Phi_2 &= \diag ( \chi^2_{(1)}(z), \cdots, \chi^{2}_{(n)}(z), 0, \cdots, 0 ) , \\
 \Phi_3 &= \diag ( \chi^3_{(1)}(z), \cdots, \chi^{3}_{(n)}(z), 0, \cdots, 0 ) .
\end{split}\label{general_less}
\end{equation}
This configuration has singularities and monodromies around those singularities.  Since the monodromies are permutations of the roots, the monodromy group $M$ is a subgroup of the symmetric group $S_{n}$, so a subgroup of SU$(n)$ and SU$(N)$. Therefore, the monodromy around a singularity can be cancelled by introducing appropriate gauge holonomy around the singularity as we did in section \ref{sec:1/4def}.  As a result, the configuration \eqref{general_less} with the appropriate gauge holonomy is a well-defined configuration. Moreover this configuration with $A_{\mu}=0$ is a solution of the equations of motion.

The scalar field configuration \eqref{general_less} and the holonomy break the gauge group SU$(N)$ to U$(1)\times$SU$(N-n)$ in the most typical case. Therefore one can introduce further holonomy in this remaining U$(1)$ around singularities. Naively this extra holonomy is parameterized by $m$ parameters $\alpha_1,\dots, \alpha_m$, where $m$ is the number of singularities.  More precisely speaking, the problem is to classify the flat connections which satisfy the conditions
\begin{enumerate}
 \item The holonomy group commutes with U$(1)\times$SU$(N-n)$.
 \item The adjoint action of the holonomy to the scalar field configuration \eqref{general_less} cancels the monodromy.
\end{enumerate}
This problem seems to be a rather non-trivial one and we will not pursue it any more in this paper. One can also introduce the theta angles for the remaining U$(1)$ gauge fields on the singularities. Naively it is parameterized by $m$ parameters $\eta_1,\dots,\eta_m$.

We define the surface operator by the path-integral with the boundary condition at the singularities such that the fields have the same singularity as the configuration \eqref{general_less} including the holonomy.

This operator preserves dilatation symmetry \eqref{dilatation} 
because $\Phi_a(z)$ are homogeneous degree $(-1)$ functions of $z^m$. 
Thus we can identify the radial direction as the time and perform the Wick rotation. We limit ourselves to these dilatation invariant surface operators in this paper.

This operator in general preserves 1/16 of the supersymmetry. This can be seen as follows. The fermion variation in the background \eqref{general_less} becomes
\begin{equation}
  \delta \psi =\sum_{a=1,2,3}\left[ \half  ( \delb_{\bar i}  \bPhi_{a} \gamma^{ \bar{i} } \gamma^{a+2} + \partial_{i}  \Phi_a \gamma^{i} \gamma^{ \overline{a+2} }  ) \epsilon (z) - (  \bPhi_{a} \gamma^{a+2}  + \Phi_{a} \gamma^{\overline{a+2}}  ) \epsilon_1\right], \label{var1/16} 
\end{equation}
where $\epsilon(z)$ is defined as \eqref{epsilon(z)}.
The variation \eqref{var1/16} vanishes if one requires the conditions for the parameters $\epsilon_{j},\ j=1,2$
\begin{equation}
 \begin{split}
&   \gamma^{\bar{1}}\gamma^{3}\epsilon_j= \gamma^{1 } \gamma^{\bar{3}}\epsilon_j 
=   \gamma^{\bar{1}}\gamma^{4}\epsilon_j= \gamma^{1 } \gamma^{\bar{4}}\epsilon_j 
=   \gamma^{\bar{1}}\gamma^{5}\epsilon_j= \gamma^{1 } \gamma^{\bar{5}}\epsilon_j 
=0,\\
&   \gamma^{\bar{2}}\gamma^{3}\epsilon_j= \gamma^{2 } \gamma^{\bar{3}}\epsilon_j 
=   \gamma^{\bar{2}}\gamma^{4}\epsilon_j= \gamma^{2 } \gamma^{\bar{4}}\epsilon_j 
=   \gamma^{\bar{2}}\gamma^{5}\epsilon_j= \gamma^{2 } \gamma^{\bar{5}}\epsilon_j 
=0.
 \end{split}
\label{cond-1/16-1}
\end{equation}
Actually, this condition \eqref{cond-1/16-1} is equivalent to the condition
\begin{equation}
 -\epsilon_j=\Gamma^{0145}\epsilon_j=\Gamma^{0167}\epsilon_j=\Gamma^{0189}\epsilon_j=\Gamma^{2345}\epsilon_j.
\end{equation}
Hence one can see that $1/16$ of the supersymmetry and $1/16$ of the super-conformal symmetry are preserved by the surface operator \eqref{general_less}.
\subsection{Gravity dual of 1/16 BPS surface operators}
Here let us consider the gravity dual of the surface operator defined in eq.\eqref{general_less}.

We propose that the gravity dual of the surface operator characterized by \eqref{general_less} will be a D3-brane wrapping a holomorphic sub-space $\Sigma_4$ defined by three holomorphic equations 
\begin{equation}
 f_a (z^{1},z^{2}, \mu\omega^1, \mu\omega^2, \mu\omega^3) = 0,  \quad ( a = 1,2,3 ),\qquad
 \mu\equiv\frac{\sqrt{\lambda}}{2\pi}.\label{sigma4}
\end{equation}
Here we use the coordinates of eq.\eqref{metric}. As we will see, this reproduces the 1/2 BPS case in \cite{Constable:2002xt,Gukov:2006jk,Gomis:2007fi} and the 1/4 BPS case in section \ref{sec:gauge theory} and section \ref{sec:gravity theory}.

To see the supersymmetry it is convenient to go to Lorentzian signature and 12 dimensional notation of appendix \ref{12d_convention}. The 4-dimensional sub-space \eqref{sigma4} is expressed as the intersection of $AdS_5\times S^5$ and the 6-dimensional sub-manifold $\Sigma_6$ in 12 dimensions.  This $\Sigma_6$ is described by the algebraic equation
\begin{equation}
 f_a (Z^{1},Z^{2}, \mu W^1, \mu W^2, \mu W^3) = 0,  \quad ( a = 1,2,3 ).\label{sigma6}
\end{equation}
Actually the supersymmetry of this class of D3-brane is checked by Kim and Lee \cite{Kim:2006he}\footnote{They consider in \cite{Kim:2006he} more general time dependent configurations of D3-brane and show that they preserve 1/16 of the supersymmetry.}.  They have shown that in general it preserves at least 1/16 of the supersymmetry.
\subsection{Examples}

In general this surface operator preserves $1/16$ of the supersymmetry.  However in some particular cases, this surface operator preserves larger amount of supersymmetry.
In this subsection we explain some examples of 1/2, 1/4, and 1/8 BPS cases.

\subsubsection{1/2 BPS surface operators}
As an example, this surface operator becomes a $1/2$ BPS surface operator when the functions $f_p,\ (p=1,2,3)$ become
\begin{equation}
 f_1=f_1(z^1,\chi^1),\qquad
f_2=\chi^2,\qquad
f_3=\chi^3. \label{general 1/2}
\end{equation}
Since $f_1$ is homogeneous function of $(z^1,\chi^1)$ when they are assigned the weights $(+1,-1)$, $f_1=0$ has $n=n_1$ roots of the form
\begin{align}
 \chi^{1}_{(i)}=\frac{\beta_i}{z^1},
\end{align}
where $\beta_i,\ (i=1,\dots,n)$ are constants. As a result one finds that the operator \eqref{general_less} with $f_p$ of eq.\eqref{general 1/2} are the 1/2 BPS operators discussed in \cite{Gukov:2006jk,Gomis:2007fi,Drukker:2008wr}.
The supersymmetry for this operator is considered in \cite{Gukov:2006jk,Gomis:2007fi,Drukker:2008wr,Constable:2002xt,Lin:2004nb,Lin:2005nh} in both gauge theory side and supergravity side. It is found that this operator is actually 1/2 BPS. We explain the supersymmetry of the probe D3-brane picture in appendix \ref{half_12d} from the 12-dimensional point of view.
\subsubsection{1/4 BPS surface operators}
\label{sec:general 1/4}
The general surface operator becomes a $1/4$ BPS surface operator when $f_a$ are written as
\begin{equation}
f_1=f_1(z^1,z^2,\chi^1),\qquad
f_2=\chi^2,\qquad
f_3=\chi^3. \label{1/4 f}
\end{equation}

A more special examples is 
\be f_1 = g(z^1, z^2) \left( \chi^1 \right)^n - \beta^n, \quad f_2 = \chi^2, \quad f_3 = \chi^3 ,
\nn \ee
where $g(z^1, z^2)$ is a degree $n$ homogeneous polynomial of $z^1, z^2$. One can factorize this polynomial and write it as $ g(z^1, z^2) = \prod_{i=1}^n ( a_i z^1 + b_i z^2) $ for constants $a_i,b_i$.  This operator is localized at planes $a_i z^1 + b_i z^2=0,\ (i=1,\dots,n)$, which are intersecting at a point $z^1 = z^2= 0$. 

Another example is that $f_1 ( z^1, z^2, \chi^1) $ is a homogeneous function when $z^1, z^2, \chi^1 $ are assigned weights $(1,1,-1)$ and $(A, B, 0 )$. For the case,  $f_1$ can be expressed as 
\begin{equation} f_1  (z^1, z^2, \chi ) = \sum_l  c_l ( (z^1)^{ -  \frac{B}{A-B} } (z^2)^{\frac{A}{A-B} }  \chi^1 )^{ l} (z^1)^{ \frac{ - B k_1 + k_2}{ A-B}  } ( z^2 )^{  \frac{ A k_1 - k_2}{A-B}  } ,    \nn 
\end{equation} 
for some constant coefficients $c_l$. Solving $f_{1} (z^1, z^2, \chi^1 )  = 0 $ amounts to solving an algebraic equation of one variable, $(z^1)^{ -  \frac{B}{A-B} } (z^2)^{\frac{A}{A-B} }  \chi^1$.  Thus the solution is in the form of  
\begin{equation}
  \chi^1_{(i)}  \sim  (z^1)^{  \frac{B}{A-B} } (z^2)^{  - \frac{A}{A-B} } . 
\nn 
\end{equation} 

The 1/4 BPS surface operator  which we considered in section \ref{sec:gauge theory} and section \ref{sec:gravity theory} is a special example that 
\begin{equation}
f_1=z^1z^2(\chi^1)^2-\beta^{2},\qquad
f_2=\chi^2,\qquad
f_3=\chi^3. \label{quarter_ftn}
\end{equation}

The supersymmetry of the operators with \eqref{1/4 f} can be seen just the same way as done in section \ref{sec:gauge SUSY} and section \ref{sec:gravity SUSY}.

The correlation function with chiral primary operators can also be calculated in the similar way to section \ref{sec:cpo gauge} and section \ref{sec:cpo gravity}.  In the gauge theory side, we can just insert the classical solution \eqref{general_less} with \eqref{1/4 f} into the form \eqref{SO(4)inv-cpo} and we get
\begin{equation} 
\frac{\langle \cO_{ \Delta , k } (\zeta)\cdot \cO_{ \Sigma} \rangle}{ \langle \cO_{ \Sigma} \rangle } =   \frac{ ( 8 \pi^2)^{ \Delta /2 } }{ \lambda^{ \Delta /2} \sqrt{ \Delta } } C_{ \Delta, k }  \sum_{i=1}^{n}
\left(\chi^1_{(i)}(\zeta)\right){}^{\frac{\Delta+k}{2}}
\overline{\left(\chi^1_{(i)}(\zeta)\right)}{}^{\frac{\Delta-k}{2}}
. 
\label{cpo:gauge general} 
\end{equation} 
On the other hand, in the gravity side, eq.\eqref{int:quarter} is still valid for the D3-brane of \eqref{sigma4} with \eqref{1/4 f}. We consider the same limit as in section \ref{sec:cpo gravity} and evaluate the integral, taking care of the branches with $\omega^1(z)=\mu^{-1}\chi^1_{(i)}(z)$. The leading term become
\begin{equation} 
\begin{split}
 &\frac{\langle \cO_{ \Delta , k } (\zeta)\cdot \cO_{ \Sigma} \rangle}{ \langle \cO_{ \Sigma} \rangle } = 
\sum_{i=1}^{n}I_{0,(i)},\\
&I_{0,(i)}\equiv  -\frac{2^{\Delta/2}}{\sqrt{\Delta}}
C_{ \Delta, k }\;\mu^{-\Delta} \left(\chi^1_{(i)}(\zeta)\right){}^{\frac{\Delta+k}{2}}
\overline{\left(\chi^1_{(i)}(\zeta)\right)}{}^{\frac{\Delta-k}{2}}
. 
\end{split}
\label{cpo:gravity general} 
\end{equation} 
This gravity result \eqref{cpo:gravity general} completely agrees with the gauge theory result \eqref{cpo:gauge general}. The next-to-leading order can be also done as before. Including this correction term, the correlation function calculated in the gravity side become
\begin{equation}
 \frac{\langle \cO_{ \Delta , k } (\zeta)\cdot \cO_{ \Sigma} \rangle}{ \langle \cO_{ \Sigma} \rangle } = 
\sum_{i=1}^{n}I_{0,(i)}\left[1+\mu^2 \frac{\Delta^2-k^2}{4(\Delta-1)\lvert \chi^{1}_{(i)}(\zeta)\rvert^4}
(\lvert\del_1 \chi^{1}_{(i)}(\zeta)\rvert^2
 +\lvert\del_2 \chi^{1}_{(i)}(\zeta)\rvert^2)\right].
\end{equation}
Because $\mu^2=\lambda/(2\pi)^2$, this series is a positive power expansion in $\lambda$. Thus we may expect that this result can be compared to the perturbative gauge theory.
\subsubsection{1/8 BPS surface operators I} 
In this paper, we mainly consider less BPS surface operators with monodromy. There are less BPS surface operators which have no monodromy and a singularity at $z^1=0$, with $\Phi_a\sim 1/z^1$. 

Let us take the functions $f$'s as $z^2$ independent.
\begin{equation}
 f_1=f_1(z^1,\chi^1,\chi^2,\chi^3),\qquad
 f_2=f_2(z^1,\chi^1,\chi^2,\chi^3),\qquad
 f_3=f_3(z^1,\chi^1,\chi^2,\chi^3).
\end{equation}
The solutions $\chi_{(i)}^{a}(z^1)$ of the algebraic equation $\{f_p=0\}$ are degree 1 functions of $z^1$, and therefore they have the form $\chi_{(i)}^{a}(z^1)=\beta_{a,i}/z^1$ with some constants $\beta_{a,i}$.
Thus the classical configuration for this class of operator is parameterized by $3\times n$ complex numbers $\beta_{a,i},\ (a=1,2,3,\ i=1,\dots,n)$ and written as
\begin{equation}
 \Phi_a=\frac{1}{z^1} \diag(\beta_{a,1},\dots,\beta_{a,n},0,\dots,0).
\label{1/8}
\end{equation}

The background filed as in \eqref{1/8} preserves $1/8$ of supersymmetry and super-conformal symmetry.  The imposition of the following conditions on $ \epsilon_i $, 
\begin{equation} 
 \gamma^1 \gamma^{ \bar{3} } \epsilon_i = \gamma^1 \gamma^{ \bar{4} } \epsilon_i = \gamma^1 \gamma^{ \bar{5} }  \epsilon_i 
=\gamma^{\bar{1}} \gamma^{3 } \epsilon_i = \gamma^{\bar{1}} \gamma^{ 4 } \epsilon_i = \gamma^{\bar{1}} \gamma^{ 5 }  \epsilon_i = 0 , \quad ( i = 0, 1), \nn \end{equation} 
or equivalently
\begin{equation}
 -\epsilon_j=\Gamma^{0145}\epsilon_i=\Gamma^{0167}\epsilon_i=\Gamma^{0189}\epsilon_i,
\end{equation}
makes the super-conformal transformation (\ref{c:var}) vanish. So this operator preserves 4 supersymmetries.

To get a $1/4$ BPS surface operator, we let $f_3 ( z^1, \chi^3) = \chi^3 $. In that case,  $-\epsilon_j=\Gamma^{0145}\epsilon_i=\Gamma^{0167}\epsilon_i$ suffices to have $ \delta \psi = 0$ for the corresponding surface operator. 

For the less BPS surface operators in this  section,  the evaluation of the classical action will be a linear sum of  that of the half BPS surface operators , which is shown to vanish. Thus the vacuum expectation value of the operators will be also $1$. 

 The gravity dual of this operator is given by $n$ disconnected D3-brane sheets
\begin{equation}
 \omega_{a}=\frac{\beta_{a,i}}{\mu z^{1}},\qquad
\mu=\frac{\sqrt{\lambda}}{2\pi},\qquad i=1,\dots,n.
\end{equation}
Actually each sheet is a 1/2 BPS configuration, and the preserved supersymmetry can be seen as in appendix \ref{half_12d}. The preserved supersymmetry by $i$-th sheet is
\begin{equation}
 \gamma_{1}\sum_{a=1,2,3}\bar{\beta}_{a,i}\gamma_{\overline{a+2}}\epsilon
=\gamma_{\bar{1}}\sum_{a=1,2,3}\beta_{a,i}\gamma_{{a+2}}\epsilon=0.
\end{equation}
Thus if we impose
\begin{equation}
 \gamma_{1}\gamma_{\bar{a}}\epsilon
=\gamma_{\bar{1}}\gamma_{a}\epsilon=0,\qquad
a=3,4,5,
\end{equation}
then the SUSY variation vanishes. As a result one can see that 1/8 of the supersymmetry is preserved.

\subsubsection{1/8 BPS surface operators II}
There is another class of 1/8 BPS surface operators expressed by the following equations.
\begin{equation}
f_1=f_1(z^1,z^2,\chi^1,\chi^2),\qquad
f_2=f_1(z^1,z^2,\chi^1,\chi^2),\qquad
f_3=\chi^3. \label{1/8 f}
\end{equation}
The preserved supersymmetry is expressed by
\begin{equation}
0= \gamma^{1}\gamma^{\bar{3}}\epsilon_{i}
= \gamma^{\bar{1}}\gamma^{3}\epsilon_{i}
= \gamma^{1}\gamma^{\bar{4}}\epsilon_{i}
= \gamma^{\bar{1}}\gamma^{4}\epsilon_{i}
= \gamma^{2}\gamma^{\bar{3}}\epsilon_{i}
= \gamma^{\bar{2}}\gamma^{3}\epsilon_{i}
= \gamma^{2}\gamma^{\bar{4}}\epsilon_{i}
= \gamma^{\bar{2}}\gamma^{4}\epsilon_{i},
\end{equation}
or equivalently
\begin{equation}
 -\epsilon_{i}
 =\Gamma^{0145}\epsilon_{i}
 =\Gamma^{0167}\epsilon_{i}
 =\Gamma^{2367}\epsilon_{i}.
\end{equation}
Therefore this operator \eqref{1/8 f} is actually 1/8 BPS.

When the equations take the special form as
\begin{equation}
f_1=f_1(z^1,\chi^1),\qquad
f_2=f_1(z^2,\chi^2),\qquad
f_3=\chi^3,
\end{equation}
then this operator preserves 1/4 of the supersymmetry expressed as
\begin{equation}
 -\epsilon_{i}
 =\Gamma^{0145}\epsilon_{i}
 =\Gamma^{2367}\epsilon_{i}.
\end{equation}

\section{Discussion}
\label{sec:discussion}
In this paper, we discuss a class of BPS surface operators.  These operators are defined by a set of algebraic equations. These operators preserve in general $1/16$ of the supersymmetries. We propose the holographic dual of those operators as configurations of probe D3-branes. We checked the supersymmetry in both the gauge theory side and the gravity side.

We took a special example of a 1/4 BPS surface operator \eqref{quarter:op} and studied it in more detail. We calculated the expectation value of this operator in both the gauge theory side and the gravity side, and found it is $1$ in both calculation. We also considered the correlation functions with local chiral operators. In the leading order at $\lambda/\beta^2\to 0$, both the calculations completely agree with each other (see \eqref{cpo:gauge} and \eqref{cpo:gravity}). We also calculated next-to-leading order contribution in the gravity side and got the result \eqref{correction}. This correction, unlike the 1/2 BPS case\cite{Drukker:2008wr}, includes space-time position dependence. This is because the spacetime dependence is not completely fixed the remaining spacetime symmetry in the 1/4 BPS case, while it is fixed by the remaining conformal symmetry in the 1/2 BPS case ($d^{-\Delta}$ behavior).

In the expectation value calculation we only consider the ``flat'' surface operator. It preserves some Q supersymmetry, and the trivial expectation value is a consequent of this supersymmetry. It will be interesting to consider the curved 1/4 surface operator and calculate the anomaly similar to ones considered in \cite{Berenstein:1998ij,Graham:1999pm,Henningson:1999xi,Gustavsson:2003hn,Gustavsson:2004gj} in 6-dimensional CFT.

In the calculation of the correlation function with chiral primary operators in the gravity side, we somehow get the positive power expansion in $\lambda$. This is because in our case $\beta$ can be large and the actual expansion parameter is $\lambda/\beta^2$. This situation is quite similar to the plane wave limit \cite{Berenstein:2002jq} in which the R-charge $J$ is large and $\lambda/J^2$ becomes the expansion parameter. It will be very interesting to calculate the next-to-leading order in the Yang-Mills theory side and see if it agrees with the result of the gravity side.

In the result \eqref{correction}, the next-to-leading term vanishes when  $\Delta=|k|$. The same thing happens in the 1/2 BPS case \cite{Drukker:2008wr}. Actually in the 1/2 BPS case it is observed that the power series terminates at a finite order for every chiral primary operator.  It is not clear whether the same thing happens in 1/4 BPS case. It will be interesting to see if this power series terminates at a finite order.

To consider the operator spectrum in 1/4 (or less BPS) surface operator as done in \cite{Constable:2002xt} is also an interesting problem. Actually the surface operator treated here preserves some supersymmetry, the index considered in \cite{Romelsberger:2005eg,Kinney:2005ej,Nakayama:2005mf,Nakayama:2006ur}
may have some interesting property.

One can also consider the correlation function with other kinds of operators, for example, Wilson loops.  The holographic dual of the Wilson loops has various descriptions: fundamental string probe, D3-brane probe, D5-brane probe and bubbling geometry \cite{Rey:1998ik,Maldacena:1998im,Drukker:2005kx,Hartnoll:2006hr,Yamaguchi:2006tq,Gomis:2006sb,Gomis:2006im,Yamaguchi:2006te,Lunin:2006xr,D'Hoker:2007fq}. It will be an interesting problem to calculate the correlation function with the Wilson loop using these descriptions.

In this paper, we only use probe D3-branes to describe the gravity dual of the less BPS surface operators. It will be a challenging problem to include the back reaction and construct the supergravity solution for these less BPS surface operators. There are several works on less BPS bubbling geometry \cite{Kim:2005ez,Donos:2006iy,Donos:2006ms,Gava:2006pu,Gauntlett:2006ns,Chen:2007du,Gauntlett:2007ph,Lunin:2008tf,Donos:2008ug,Donos:2008hd}. One may find the solution of 1/4 BPS surface operator in these backgrounds, or doubly Wick rotated ones.

Finally one can also consider the surface operators in 4-dimensional $\N=1$ super-conformal field theory and their gravity dual in the type IIB string theory on AdS${}_5\times$(Sasaki-Einstein). In the CFT side, the surface operator can be considered by imposing boundary condition for the complex scalar fields in the chiral superfield. In the gravity side, it will correspond to some configuration of D3-branes. This configuration is described by a 6-dimensional homogeneous holomorphic submanifold in $\Cb^{1,2}\times $(Calabi-Yau 3-fold cone) in 12-dimensional picture.

\subsection*{Acknowledgments}
We would like to thank Nadav Drukker for careful reading of the manuscript and helpful comments. We are also grateful to Soo-Jong Rey and Takao Suyama for useful discussions.  A part of this work was done during the YITP workshop YITP-W-08-04 on ``Development of Quantum Field Theory and String Theory,'' (Jul. 27- Aug.\ 1, 2008) and ``Summer Institute 2008'' at Yamanashi (Aug.\ 3-13, 2008). 
This work was supported in part by KOFST BP Korea Program, KRF-2005-084-C00003, EU FP6 Marie Curie Research and Training Networks MRTN-CT-2004-512194 and HPRN-CT-2006-035863 through MOST/KICOS.

\appendix

\section{Conventions for Gauge Theory}  \label{gauge}
The action of $ \N=4 $ SYM  on 4 dimensions can be written as $\N=1$ SYM in 10 dimensions:
\begin{equation} 
S = \frac{1}{g^2} \int d^4 x \tr \left[ \frac{1}{2} F_{MN} F^{MN} - i \bar{ \psi} \Gamma^M D_M \psi \right],  \label{action}
\end{equation} 
where  $M,N= 0, 1, \cdots, 9 $ , $ \Gamma^M$ is the gamma matrix in 10 dimensions.  As we consider the theory in 4 dimensions, 
 for $ M = 4, 5, \cdots, 9$ , 
$ \partial_M = 0 $  and the gauge fields $A_M$ become six real scalars $ \phi_1, \cdots \phi_6$ . 
We define three complex scalars as follows : 
\begin{equation}
 \Phi_n \equiv \phi_{2n-1} + i \phi_{2n},  \quad n=1, 2, 3. 
\end{equation}
We define complex gamma matrices as follows: 
\begin{equation} 
\begin{cases}
 \gamma^A \equiv & \Gamma^{2A-2} + i \Gamma^{ 2A -1} ,   \\ 
\gamma^{ \bar{A} }  \equiv & \Gamma^{ 2A -2 } - i \Gamma^{ 2A -1 } , \quad ( A = 1, 2, \cdots, 5 ). \end{cases}
\label{10d:gamma}
\end{equation} 

The supersymmetry and super-conformal symmetry transformation are given as  
\begin{align}
\delta \psi =& \left( \half F_{ \mu \nu} \Gamma^{ \mu \nu} + D_{ \mu} \phi_I \Gamma^{ \mu I} - \frac{i}{2} [ \phi_I , \phi_J ] \Gamma^{IJ} \right) \epsilon(x)- 2 \phi_I \Gamma^I \epsilon_1,  \label{var:fermion}  \\ 
\delta A_M =&  - i \bar{ \psi} \Gamma_M \epsilon(x) ,   \nn 
\end{align}
where 
\begin{equation} \epsilon(x) = \epsilon_0 + x^{ \mu} \Gamma^{ \mu} \epsilon_1 , \nn \end{equation} 
for 16 real components constant Majorana-Weyl spinors $ \epsilon_i$, $i = 0, 1$.

Conventions for the complex coordinates are 
\begin{align}
 & z^1 = x^0 + i x^1, \quad z^2 = x^2 + i x^3 , \nn \\
& ds^2  = dz^1 d \bar{z}^1 + dz^2 d \bar{z}^2 = 2 \eta_{A \bar{B} } dz^A d \bar{z}^{ \bar{B}}, \qquad
\eta_{ A \bar{B} }  = \begin{pmatrix}  \half & 0  \\  0 & \half  \end{pmatrix}, \quad \nn \\
& \del_{m}\equiv \frac{\del}{\del z^m} = \half \left( \frac{\del}{\del x^{2m-2}} - i \frac{\del}{\del x^{2m-1}} \right), \qquad\delb_{\bar{m}}\equiv \frac{\del}{\del \bar{z}^{\bar m}} = \half \left( \frac{\del}{\del x^{2m-2}} + i \frac{\del}{\del x^{2m-1}} \right), \quad \nn  
\end{align}

We define the measure as 
\begin{equation} dz^1 d \bar{z}^{ \bar{1} } \equiv d^2 z^1 \equiv 2 dx^0 dx^1 = 2 r_1 dr_1 d \phi_1 , \quad dz^2 d \bar{z}^{ \bar{2} } \equiv d^2 z^2 \equiv 2 dx^2 dx^3 = 2 r_2 dr_1 d \theta_2.  \label{measure} 
\end{equation} 

\section{Kappa Symmetry Projection} \label{kappa}

$ \Gamma$ can be defined as follows for a $D3$ brane without electro-magnetic flux: 
\begin{align}
 \Gamma =& \frac{1}{ \sqrt{ - \det G} } \Gamma_{A_0 A_1 A_2 A_3 } E^{A_0}_{M_0} E^{A_1}_{M_1}   E^{A_2}_{M_2}  E^{A_3}_{M_3} \frac{ \partial X^{M_0} }{ \partial \xi^0 }  \frac{ \partial X^{M_1} }{ \partial \xi^1 }  \frac{ \partial X^{M_2} }{ \partial \xi^2 }  \frac{ \partial X^{M_3} }{ \partial \xi^3 }  \label{projection}   \\
=& \frac{1}{ \sqrt{- \det G }}  \frac{ \epsilon^{ \mu \nu \rho \sigma} }{ 4 ! } \Gamma_{A_0} \Gamma_{A_1} \Gamma_{A_2} \Gamma_{A_3}  E_{ M^0}^{A_0}  E^{A_1}_{M_1}   E^{A_2}_{M_2}  E^{A_3}_{M_3} \frac{ \partial X^{M_0} }{ \partial \xi^\mu }  \frac{ \partial X^{M_1} }{ \partial \xi^{ \nu} } \frac{ \partial X^{M_2} }{ \partial \xi^{ \rho} }  \frac{ \partial X^{M_3} }{ \partial \xi^{ \sigma} }, \label{projection:1}  
\end{align}
where $G$ is the induced metric on the world volume, $\xi^\mu$  world-volume coordinate, $X^{ M_i}$ space-time coordinates, and $A_i$ flat directions. We choose a convention $ \epsilon^{0123}=1$. 
If we can set $ \xi^0 = t$, 
\begin{equation}
 \Gamma = \frac{1}{ \sqrt{  \det G_{space} } }    \Gamma_0  \Gamma_{ A_1 A_2 A_3 } E^{A_1}_{M_1}   E^{A_2}_{M_2}  E^{A_3}_{M_3}   \frac{ \partial X^{M_1} }{ \partial \xi^1 }  \frac{ \partial X^{M_2} }{ \partial \xi^2 }  \frac{ \partial X^{M_3} }{ \partial \xi^3 }.   \label{gamma}   
\end{equation} 

For the type IIB theory, the number of preserved supersymmetries by the Dp-brane is the number of Killing spinors satisfying
\begin{equation}  \Gamma K^{ \frac{p+1}{2} } I \epsilon = \epsilon,  \nn \end{equation} 
where $ I \epsilon =  i \epsilon $ and $ K \epsilon = \epsilon^{ \ast } $.  For the D3-brane, the condition becomes
\begin{equation} i \Gamma \epsilon =  \epsilon  . \label{D3}
\end{equation} 

\section{Conventions for 12-dimensional Space} \label{12d_convention}  
Let us consider $ \Rb^{ 2, 4} \times \Rb^6$ with coordinates $( X^{-1}, X^0, X^1,\dots, X^4, Y^1, \dots, Y^6 )$. We define complex coordinates as 
\begin{align}
Z^A =&  X^{2A-1} + i X^{2A} , \quad ( A =  0, 1, 2),   \nn \\
W^A =& Y^{2A-1} + i Y^{2A}  , \quad ( A= 1,2,3) .
\nn  
\end{align}

The 12-dimensional metric is 
\begin{align}
  ds^2_{12}  =&  -  dZ^0 d \bar{Z}^0 + \sum_{A=1}^2 dZ^A d \bar{Z}^{ \bar{ A}}  + \sum_{A=1}^3 d W^A d \bar{W}^{\bar{A}}  \nn \\
 =& 2 \eta_{A \bar{B} } dZ^A d\bar{Z}^{ \bar{B} }, \quad (A, B =0, 1,2 , \cdots, 5 ),   \nn
\end{align}
where $ ( Z^4, Z^5, Z^6)  \equiv ( W^1, W^2, W^3) $ in the second line.

Define the complex gamma matrices as : 
\begin{equation} 
\begin{split}
 \gamma^A  \equiv & \Gamma^{2A-1} + i \Gamma^{2A}, \\
 \gamma^{ \bar{A}}  \equiv & \Gamma^{2A -1} - i \Gamma^{2A} ,  \quad  (A=0, \cdots, 5) ,
 \end{split}
 \label{cplx_gamma} 
\end{equation} 
where $ \Gamma^{A}$ are 12-dimensional gamma matrices not to be confused with 10 dimensional ones in (\ref{10d:gamma}). $ \gamma^A$ satisfy
\begin{equation} \{ \gamma^A, \gamma^{ \bar{B}} \} = 2 \eta^{ A \bar{B}}. \nn \end{equation} 
Gamma matrices with a lower index are defined by: 
\begin{equation}  \gamma_A  \equiv \eta_{A \bar{B}} \gamma^{ \bar{B} }, \quad \gamma_{ \bar{A} } \equiv \eta_{ \bar{A}  B  } \gamma^B .
\nn 
\end{equation} 
We define 
\begin{equation}
  \gamma_{ \bar{A} B} \equiv \gamma_{ [ \bar{A}, B ] } \equiv \half \left( \gamma_{ \bar{A} } \gamma_B - \gamma_B \gamma_{ \bar{A} } \right) .
\nn 
\end{equation} 
A useful identity is : 
\begin{equation} \gamma_{ \bar{A}  B  } \gamma_{  \bar{C}  D  } =  \gamma_{  \bar{C} D } \gamma_{  \bar{A}  B  } + 2 \eta_{ \bar{A} D } \gamma_{  B  \bar{C}  } + 2 \eta_{ \bar{B} C } \gamma_{  \bar{A}  D  } .
\nn 
\end{equation} 

\section{The gravity dual of the half BPS surface operator}
\subsection{The half BPS D3-brane configuration in 12 dimensions}
\label{half_12d} 
We consider a 6-dimensional hyperspace defined by holomorphic functions
\begin{equation} \tilde{f}( Z^1,  W^1)=0, \quad W^2 =0,\quad W^3 = 0, \nn 
\end{equation} 
where 
$\tilde{f}( Z^1, W^1)=f(Z^1,(2\pi/\sqrt{\lambda})W^1)$ is a homogeneous function of $Z^1$ and $W^1$ when they are assigned weights $ (+1, -1)$. 
The normal vectors can be chosen as 
\begin{align}
  E_{r_1} =& \frac{ 1}{ \sqrt{ 1 + \ab } } \left( Z^0 \partial_0 + Z^2 \partial_2 + c.c \right), \label{vec:r1} \\
 E_{r_2 }=&   \frac{1}{ \sqrt{ 1 + \ab} } \left( Z^1 \partial_1 - Z^3 \partial_3 + c.c   \right) .
\nn 
\end{align}
The tangent vectors can be specified as follows : 
\begin{align}
  E_z =& \frac{1}{ \sqrt{  1+ \vert Z^0 \vert^2 } } \left( \bar{Z}^{ \bar{2} } \frac{ \partial}{ \partial Z^0 } + \bar{Z}^0 \frac{ \partial}{ \partial Z^2 } \right),  \quad E_{ \bar{z} } = E_z^{ \ast},  \label{vec:z}  \\ 
E_{ \pm } =& \frac{1}{2} \left(   \pm I \cdot E_{r_1} + I \cdot E_{r_2}  \right).   \nn  
\end{align}

Impose the half BPS condition 
\begin{equation}
 \gamma_{ 1  \bar{1 }  } \epsilon =  \gamma_{ 3  \bar{3} } \epsilon .
\label{halfbps:gravity}
\end{equation} 
If multiplied by $ \gamma_{ \bar{1}} \gamma_{ 3}$ or $ \gamma_{ \bar{3}} \gamma_{ 1} $, it implies 
\begin{equation} 
\gamma_{ \bar{1}} \gamma_{ 3 } \epsilon = \gamma_{ \bar{3}} \gamma_{ 1} \epsilon = 0.  \nn 
\end{equation} 
The condition reduces the left hand side of (\ref{12d_projection}) to $i \epsilon $ , which shows that it preserves half of the supersymmetries. 

\subsection{Correlation function of the half BPS surface operator and a chiral primary operator}
The correlation functions of CPO's with the half BPS surface operator described by the D3-brane solution $ \omega(z^1) = \kappa/ z^1 $  can be found in \cite{Drukker:2008wr}. Here we write down the result in our notation.
\begin{align} 
&\frac{ \delta S_{D3} }{ \delta s_0^{ \Delta , k }(\zeta) } = - 2 \TD  \Delta C_{ \Delta , k } c( \Delta ) \kappa^{ - \Delta } \int d^4 z \frac{  \omega^{  - \frac{ \Delta - k }{2}}(z^1) \bar{ \omega}^{ -  \frac{ \Delta+ k}{2}} (\bar{z}^{ \bar{1} })  }{  L^{ \Delta +2} }  \frac{   \vert \zeta^1 \del_1 \omega (z^1) \vert^2  }{\vert \omega \vert^2   },
\label{int} \\
&L = L ( \omega(z), z^i - \zeta^i )  \equiv  \left(  \vert \omega(z) \vert^{-2} + \sum_{m=1,2} \vert z^m - \zeta^{m} \vert^2 \right)  . 
\nn \end{align} 
In the leading order of large $\kappa$, we can replace $ \omega(z) , \del_m \omega(z) $ with $ \omega( \zeta) , \del_m \omega( \zeta)  $.  We let $ \zeta^1 = d_1 e^{ i \phi_1^{ \prime} } , \zeta^2 = 0 $.  Integrating out the $dz^2 d \bar{z}^2$ then using (\ref{integration:1}) leads to 
\begin{equation}
 \frac{\langle \cO_{ \Delta , k } \cdot \cO_{ \Sigma} \rangle}{\langle \cO_{ \Sigma} \rangle} = - \frac{ C_{ \Delta , k } }{ \sqrt{ \Delta } } 2^{ \Delta /2} e^{ - i k \phi_1^{ \prime} } \left( \frac{ \kappa}{d_1} \right)^{ \Delta }.  \label{half:leading} 
\end{equation} 
The result of the above integral is compatible with eq.(3.43) in \cite{Drukker:2008wr}, if we replace  $ \kappa $  with $ \sinh u_0 $.  Since the result in \cite{Drukker:2008wr} needs not assume the limit $\kappa \gg 1$,  it deviates at the sub-leading terms of $\kappa$ from (\ref{half:leading}).

For the next to the leading order of (\ref{int}), we first integrate out $dz^2 d \bar{z}^2$ then expand $ \omega(z^1) = \omega( \zeta^1)+ v^1 \del_1 \omega ( \zeta^1) + \cdots $. The integration up to the quadratic order of $v \equiv v^1 $ is proportional to 
\begin{equation}
\int d^2 v \frac{1}{ K^{ \Delta +1}  } \left( 1 + ( \frac{ \Delta^2 - k^2}{4} - \Delta +1 ) \frac{ \vert v \vert^2 }{ \vert \zeta \vert^2 }  + \frac{( \Delta +1)( - \Delta +1) }{K }   \frac{ \vert v \vert^2}{ \vert \zeta \vert^4} + \frac{( \Delta +1)( \Delta +2)}{ K^2}  \frac{ \vert v \vert^2}{ \vert \zeta \vert^6} + \cdots \right),  \nn 
\end{equation} 
where $K \equiv \vert \omega ( \zeta ) \vert^{-2} + \vert v \vert^2 $. The result is : 
\begin{equation}
\frac{\langle \cO_{ \Delta , k } \cdot \cO_{ \Sigma} \rangle}{\langle \cO_{ \Sigma} \rangle}
 = I_0^{(1/2)} \left( 1 + \kappa^{-2} \frac{ \Delta^2 - k^2}{ 4 ( \Delta -1) } + \cdots \right),  \nn 
\end{equation} 
 where $I_0^{(1/2)}$ is the leading order result given in  (\ref{half:leading}). 
This result is consistent with the result of  \cite{Drukker:2008wr}. 

\section{Correlation function with chiral primary operators}
\subsection{Spherical Harmonics} \label{spherical}
The $SO(4)$ invariant spherical harmonics are \cite{Drukker:2008wr,Skenderis:2007yb},
\begin{equation} Y^{ \Delta , k } ( \vartheta , \theta_1 ) = c_{ \Delta , k } y^{ \Delta , k} ( \vartheta ) e^{ i k \theta_1} ,
\label{harmonics}
\end{equation} 
where $y^{ \Delta, k }$ is related to the hyper-geometric function as follows: 
\begin{equation} y^{ \Delta , k} ( \vartheta ) = \sin^{ \vert k \vert } \vartheta \phantom{1}_2 F_1 \left( - \half ( \Delta -  \vert k \vert ) , 2 + \half ( \Delta + \vert k \vert ) , 1 + \vert k \vert ; \sin^2 \vartheta \right) . 
\nn 
\end{equation} 
$ C^{ \Delta , k }_{ i_1, \cdots, i_{ \Delta } }$ is defined as 
\begin{equation}
  C^{ \Delta , k }_{ i_1, \cdots, i_{ \Delta } } x^{i_1} \cdots x^{ i _{ \Delta} } = Y^{ \Delta , k } ( \vartheta , \theta_1 ) \label{def:C} , 
\end{equation} 
where $x^i$ are parameterized by the coordinates of $S^5$ as follows: 
\ba
x^1 &=& \sin \vartheta \cos \theta_1, \quad x^2 = \sin \vartheta \sin \theta_1, \nn \\
x^i &=& \Theta_i  \cos \vartheta , \quad ( i = 3,4,5,6), \quad  \mbox{for  } \sum_i \Theta_i^2 = 1 .  \nn 
 \ea
We normalize $c_{ \Delta , k }$ in (\ref{harmonics}) such that 
\begin{equation} 
\int_{S^5}  ( Y^{ \Delta_1, k_1 } )^{ \ast }  Y^{ \Delta_2, k_2}  = \frac{ \pi^3}{ 2^{ \Delta-1} ( \Delta +1) ( \Delta +2) } \delta^{ \Delta_1 , \Delta_2 }  \delta^{ k_1 , k_2 } , \nn 
\end{equation} 
or equivalently, 
\begin{equation} 
   2\pi^2  \int_0^{ \frac{ \pi}{2} } d \vartheta     \int_0^{ 2 \pi} d \theta_1  \cos^3 \vartheta \sin \vartheta  Y^{  * \Delta_1, k_1}( \vartheta , \theta_1 )  Y^{ \Delta_2 , k_2} ( \vartheta , \theta_1 )      = \frac{ \pi^3}{ 2^{ \Delta-1} ( \Delta+1)( \Delta+2) } \delta^{ \Delta_1, \Delta_2} \delta^{ k_1  , k_2} . 
   \nn \end{equation} 
We define $C_{ \Delta , k } $ as
\begin{equation}  Y^{ \Delta , k} ( \vartheta = \frac{ \pi}{2} , \theta_1 ) =  C_{ \Delta , k }  e^{ i k \theta_1 } . \label{def:cdk} 
\end{equation} 

\subsection{Some Useful formulas}
For constant $y$, the following holds: 
\begin{align}
 \int_{\Rb^{D}} d^D x \frac{1}{(x^2 + y^2)^{ \alpha} }=& \frac{ \pi^{ D/2} \Gamma( - D /2 + \alpha ) }{ \Gamma ( \alpha ) } y^{ D - 2 \alpha } , \label{integration:1} \\
\int_{\Rb^{D}} d^D x \frac{ x^a x^b }{ (x^2 + y^2 )^{ \alpha} }=& \delta^{ab} \frac{ \pi^{ D/2} \Gamma( - D/2 -1 + \alpha ) }{ 2 \Gamma ( \alpha ) } y^{D + 2 - 2 \alpha } .  \label{integration:2}
\end{align}

\providecommand{\href}[2]{#2}\begingroup\raggedright\endgroup

\end{document}